\documentclass[pra,twocolumn,floatfix,a4paper,superscriptaddress]{revtex4}
\usepackage{bm,color,graphicx,amsmath,txfonts}
\usepackage{float}
\usepackage[colorlinks, citecolor=blue,linkcolor=blue]{hyperref}

\begin{document}
	
\title{Barnett Effect-Induced Nonreciprocal Entanglement and Gaussian interferometric power in Magnomechanics with Optical Parametric Amplifier}

\author{Noura Chabar} 
\affiliation{Laboratory of High Energy Physics and Theoretical Physics, Department of Physics, Faculty of Sciences, Ibnou Zohr University, Agadir, Morocco} 

\author{M. Amghar} 
\affiliation{Laboratory of High Energy Physics and Theoretical Physics, Department of Physics, Faculty of Sciences, Ibnou Zohr University, Agadir, Morocco} 

\author{Shakir Ullah}
\affiliation{Department of Physics, Quaid-i-Azam University, Islamabad 45320, Pakistan}

\author{Mohamed Amazioug} \email{m.amazioug@uiz.ac.ma (Corresponding author)}
\affiliation{Laboratory of High Energy Physics and Theoretical Physics, Department of Physics, Faculty of Sciences, Ibnou Zohr University, Agadir, Morocco}
\author{Kottakkaran Sooppy Nisar} 
\affiliation{Department of Mathematics, College of Science and Humanities in Al-Kharj, Prince Sattam bin Abdulaziz University, Al-Kharj 11942, Saudi Arabia}
\affiliation{Hourani Center for Applied Scientific Research, Al-Ahliyya Amman University, Amman, Jordan}

\author{Mohammed Zakarya} 
\affiliation{Department of Mathematics, College of Science, King Khalid University, P.O. Box 9004, Abha 61413, Saudi Arabia}

\author{Gamal M. Ismail} 
\affiliation{Department of Mathematics, Faculty of Science, Islamic University of Madinah, Madinah 42351, Saudi Arabia}

\author{Abdel-Haleem Abdel-Aty} 
\affiliation{Department of Physics, College of Sciences, University of Bisha, Bisha 61922, Saudi Arabia}

\begin{abstract}
Nonreciprocity is a powerful tool in quantum technologies. It allows signals to be sent in one direction but not the other. In this article, we propose a method for achieving non-reciprocal entanglement and Gaussian interferometric power (GIP) via the Barnett effect. The YIG is coupled to a microwave cavity that interacts with an optical parametric amplifier (OPA). Due to the Barnett effect, giant nonreciprocal entanglement can emerge. By fine-tuning the cavity detuning, the GIP can  exhibits nonreciprocal behavior. All entanglements with ideal nonreciprocity can be achieved by tuning the photon frequency detuning, appropriately choosing the cavity-magnon coupling regime, the nonlinear gain, and the phase shift of the OPA. Interestingly, the amount of entanglement nonreciprocity and its resilience to thermal occupation are remarkably enhanced by increasing the gain of the OPA. This nonreciprocity can be significantly enhanced even at relatively high temperatures. Our research offers a pathway for the realization of nonreciprocal single-phonon devices, with potential applications in quantum information processing and quantum communication. This proposed scheme could pave the way for the development of novel nonreciprocal devices that remain robust under thermal fluctuations.		
\\

\end{abstract}
\date{\today}
\maketitle

\section{Introduction}

Hybrid magnonic devices, which leverage collective spin excitations (magnons) in ferrimagnetic crystals such as yttrium iron garnet (YIG) spheres known for their high spin density, long coherence time, low damping rate, and rich magnonic nonlinearities, offer a promising solid-state platform for implementing multifunctional quantum operations
 \cite{1,2,3,4,5}.  In these devices, magnons exhibit exceptional capability to interact with various quantum elements \cite{6,7}, enabling the development of applications including ternary logic gates \cite{19}, nonclassical states \cite{20}, non-Hermitian physics \cite{24}, and precision quantum measurement techniques \cite{32}.  
In particular, magnons can couple to mechanical vibrations (phonons) via magnetostrictive forces \cite{13}, a field known as cavity magnomechanics (CMM). This coupling has enabled the investigation of numerous phenomena, such as the creation of entangled and squeezed states \cite{36}, ultraslow light propagation \cite{39}, ultrasensitive sensing \cite{41}, and quantum state transfer, storage, and retrieval \cite{44}.  
Recently, experiments with highly tunable CMM devices have demonstrated notable achievements, including magnomechanically induced transparency and absorption \cite{13}, the generation of magnonic frequency combs \cite{14}, mechanical bistability \cite{15}, and microwave-to-optical signal conversion \cite{16}.

Recently, Girolami et al \cite{gip1} established a general quantitative relationship between the interferometric power of a bipartite state and the minimum precision guaranteed when estimating a parameter encoded through a unitary dynamics on one subsystem. Subsequently, Adesso  \cite{gip} advanced this line of research by deriving a closed formula for the GIP  in continuous-variable systems. The GIP for  separable or entangled Gaussian state can have a nonzero value \cite{z17}.

Nonreciprocity, which enables unidirectional transport by breaking reciprocal symmetries, plays a crucial role in many practical applications \cite{46}, including enhanced coordination in biological cell migration \cite{47}, improved self-organization in motor-active materials \cite{48}, and the realization of invisible sensing and noise-free information processing \cite{49}.  
Recent years have seen rapid progress in the development of magnetic-free nonreciprocal optical devices, leveraging nonlinearities, chiral interactions \cite{50}, atomic gases \cite{55}, optomechanical systems \cite{57}, and the relativistic Sagnac effect \cite{61}.  
Among these advances, nonreciprocal quantum entanglement has attracted particular attention for its potential to protect and enhance quantum correlations \cite{70}. Quantum entanglement, characterized by nonclassical correlations between separate quantum systems \cite{79}, serves as a fundamental resource for quantum information processing and quantum networks \cite{81,z4,z5}.  
Entanglement has been demonstrated across a wide variety of platforms, including photons \cite{84}, atoms \cite{86}, ions \cite{88}, and cavity optomechanical systems \cite{89}.  
The concept of nonreciprocal entanglement was first introduced in spinning resonators\cite{70}, where the system’s time-reversal symmetry is broken via an irreversible refractive index difference between clockwise and counterclockwise modes. This mechanism enables directional entanglement of photons and phonons, while strongly suppressing it in the opposite direction.  
Further studies have explored nonreciprocal entanglement using the Sagnac effect \cite{72}, magnon Kerr nonlinearities \cite{75}, and chiral couplings \cite{77}.  Nonreciprocal quantum control can be found also in phenomena such as photon  and magnon  blockade \cite{z2,z9,107}.

The enhancement of entanglement through parametric amplification has been explored extensively in various hybrid optomechanical–magnetic systems \cite{PRA,pla}. The inclusion of an optical parametric amplifier (OPA) has also attracted considerable interest in studies of quantum transparency \cite{z7}. Meanwhile, coherent feedback (CF) has emerged as a powerful tool, drawing significant attention for its ability to maintain system stability and strengthen entanglement \cite{z3,z1,z10}. The Barnett effect, where the rotation of a magnetic object induces alignment or magnetization, has been observed in ferromagnetic insulators \cite{97} and nuclear spin systems \cite{102}. It has enabled studies on remote magnetization switching \cite{104}, rotational vacuum friction \cite{105}, and the detection of angular momentum compensation points \cite{106}.  
A recent study has combined feedback control with the Barnett effect to manipulate nonreciprocal entanglement in a magnomechanical system \cite{n1}.\\

In this work, we propose a novel scheme to realize and control nonreciprocal bipartite  entanglement through the Barnett effect and OPA in a CMM setup . The rotation of the YIG sphere induces a Barnett frequency shift in the magnon mode, opening new possibilities for manipulating entanglement in a nonreciprocal fashion \cite{H1,77}.  Giant nonreciprocal entanglement can emerge.  All entanglements with ideal nonreciprocity can be achieved  via tuning the photon frequency detuning, appropriately choosing the cavity-magnon coupling regime and  nonlinear gain and the phase shift of OPA. Interestingly, the amount of entanglement nonreciprocity and her resilience  against magnon thermal occupation is remarkably enhanced by increasing the gain of OPA. That non-reciprocity can be significantly enhanced even at relatively high temperatures. By fine-tuning the cavity detuning and the Barnett shift frequency, we can generate a strongly non-reciprocal amount of GIP. Our scheme could pave the way for the development of novel non-reciprocal devices that remain robust under thermal fluctuations.

\section{Model and Hamiltonian}
\begin{figure}[htpb]
	\centering  
   {\includegraphics[width=0.48\textwidth]{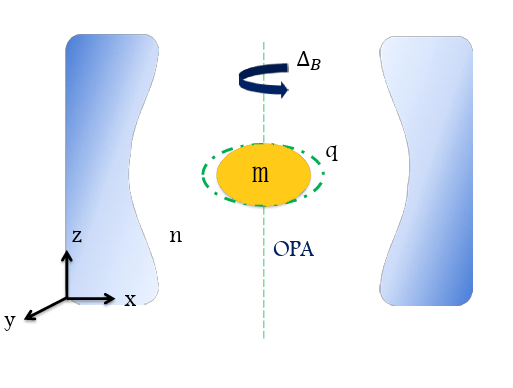}}  		 
	\caption{ (a) Schematic representation of the CMM system, consisting of a microwave cavity coupled to a YIG sphere and equipped with an OPA. The YIG sphere, which supports both a magnon mode $m$ and a phonon mode $b$, is magnetized by an external bias magnetic field $H_0$. When the sphere rotates at an angular frequency $\Delta_B$, it generates an emergent magnetic field $H_B$, leading to a frequency shift in the magnon mode.
	 }
	\label{F1A}
\end{figure}
As depicted in Fig. \ref{F1A}(a), we consider a cavity magnomechanical (CMM) system composed of a microwave cavity with frequency $\omega_n$ and a YIG sphere that supports both a magnon mode at frequency $\omega_m$, and a phonon mode at frequency $\omega_b$. The photon-magnon and magnon-phonon couplings arise from magnetic dipole and magnetostrictive interactions, respectively, and have been experimentally demonstrated in Refs \cite{6,13}. The magnetization of the YIG sphere is guaranteed by applying a bias magnetic field $H_0$ along the $z$ axis to the YIG samples. The resonance frequency of the magnon mode depends on the external magnetic field $\mathbf{H}_0$ and the gyromagnetic ratio $\gamma$, with $\omega_m = \gamma \mathbf{H}_0$, where $\gamma = 2\pi \times 28\, \mathrm{GHz/T}$. The YIG sphere is assumed to rotate about the z axis with angular frequency $\Delta_B$. According to the Barnett effect \cite{97}, this rotation gives rise to an effective magnetic field $\mathbf{H}_B = \Delta_B / \gamma$. As a result, the magnon mode undergoes a Barnett-induced frequency shift proportional to $\Delta_B$. Importantly, this angular frequency and thus the frequency shift can be affected by rotational fluctuations and thermal noise.
The magnon frequency is shifted from $\omega_m$ to $\omega_m + \Delta_B$ as a result of the Barnett effect, which arises from the conservation of angular momentum. Specifically, the Barnett-induced frequency shift of the magnon mode in the rotating YIG sphere can be tuned from positive ($\Delta_B > 0$) to negative ($\Delta_B < 0$) by reversing the direction of the applied magnetic field along the +z or -z axis.

In a frame rotating at the drive frequency $\omega_l$, the system Hamiltonian is given by ($\hbar = 1$):
\begin{equation}
	\begin{aligned}
		\hat{H}= & \Delta_n {n}^{\dagger} {n}+\left(\Delta_m+\Delta_B\right) {m}^{\dagger} {m}+\frac{\omega_r}{2}\left(\hat{q}^2+\hat{p}^2\right)+g {m}^{\dagger} \hat{m} \hat{q} 
		+  \\ & \mathcal{J} \left({n}^{\dagger} {m}+{m}^{\dagger} {n}\right)+i \varepsilon_l \left({m}^{\dagger}+{m}\right)+  
		i \chi \left(e^{i \beta} n^{\dagger 2}-e^{-i \beta} n^2  \right),
	\end{aligned}
\end{equation}

where $\Delta_{n, d}=\omega_{n, m}-\omega_l $.  $ {n}$ and ${m}$ are the annihilation operators of the photon and magnon modes, respectively.
The operators $q$ and $p$ represent the dimensionless position and momentum quadratures of the phonon mode. The symbol $\mathcal{J}(g)$ denotes the coupling rate between magnon and photon ( magnon and phonon). The drive strength is given by $\varepsilon_l = \gamma \sqrt{5N} B_0 / 4$, where $B_0$ is the field amplitude, $\omega_l$ the drive frequency, and $N = \rho V$ the total number of spins, with spin density $\rho = 4.22 \times 10^{27} \, \mathrm{m}^{-3}$ and $V$ the volume of the YIG sphere \cite{13}. Additionally, $\chi$ denotes the nonlinear gain of the optical parametric amplifier (OPA), describing the coupling between the cavity field and the OPA, with an associated phase shift $\beta$.
 
\section{Quantum Langevin equation and the steady state solution }
\label{P3}
By incorporating dissipation and noise terms phenomenologically, we obtain the quantum Langevin equations (QLEs) for the system.

\begin{equation}
\begin{aligned}
	\dot{{m}} & =-\left(i \Delta_m+i \Delta_B+\kappa_m\right) {m}-i \mathcal{J} n-i g {m} \hat{q}+\varepsilon_l+\\
	&\sqrt{2 \kappa_m} {m}^{\text {in }}, \\
	\dot{\hat{q}} & =\omega_b \hat{p}, \quad \dot{\hat{p}}=-\omega_b \hat{q}-\gamma_b \hat{p}-g {m}^{\dagger} {m}+\hat{\psi}, \\
	\dot{{n}} & =-\left(i \Delta_n+\kappa_n\right) {n}-i \mathcal{J} {m}+  2 \chi e^{i\beta} n^{\dagger}+\sqrt{2 \kappa_n} {n}^{\text {in }},
\end{aligned}
\label{3E}
\end{equation}
 $\kappa_n$ and $\kappa_m$ represent the dissipation rates of the cavity and magnon modes, respectively, while $\gamma_b$ denotes the mechanical damping rate. The operators ${m}^{\text{in}}$ and ${n}^{\text{in}}$ represent the zero-mean input vacuum noise for the microwave cavity and magnon modes, respectively, and are defined by the following correlation functions.
\begin{equation}
	\begin{aligned}
		& \langle n^{\text{in}\,\dagger}(t)\, n^{\text{in}}(t^{\prime}) \rangle 
		= \langle m^{\text{in}\,\dagger}(t)\, m^{\text{in}}(t^{\prime}) \rangle = 0, \\
		& \langle n^{\text{in}}(t)\, n^{\text{in}\,\dagger}(t^{\prime}) \rangle 
		= \langle m^{\text{in}}(t)\, m^{\text{in}\,\dagger}(t^{\prime}) \rangle 
		= \delta(t - t^{\prime}).
	\end{aligned}
\end{equation}

The operator $\hat{\psi}$ represents the Brownian noise acting on the mechanical mode, originating from its coupling to the thermal environment, and is characterized by the following correlation function.
$$
\left\langle\hat{\psi}(t) \hat{\psi}\left(t^{\prime}\right)\right\rangle=\frac{\gamma_b}{\omega_b} \int \frac{m \omega}{2 \pi} \mathrm{e}^{-\mathrm{i} \omega\left(t-t^{\prime}\right)}\left[\operatorname{coth}\left(\frac{\hbar \omega}{2 k_{\mathrm{B}} T}\right)+1\right],
$$

where $k_{\mathrm{B}}$ denotes the Boltzmann constant and $T$ is the bath temperature. Within the constraints of the high quality factor of the mechanical mode $\left(Q_b=\omega_b / \gamma_b \gg 1\right), \hat{\psi}(t)$ becomes delta correlated \cite{108}: $\left\langle\hat{\psi}^{\mathrm{in}}(t) \hat{\psi}^{\mathrm{in}}\left(t^{\prime}\right)+\hat{\psi}^{\mathrm{in}}\left(t^{\prime}\right) \hat{\psi}^{\mathrm{in}}(t)\right\rangle / 2 \simeq \gamma_b\left(2 N_b+\right.$ 1) $\delta\left(t-t^{\prime}\right)$, where $N_b=\left[\exp \left(\hbar \omega_b / k_B T\right)-1\right]^{-1}$ is the equilibrium mean thermal phonon number. Under the condition of a strong driving field, each operator can be decomposed as $\hat{o}=o_s+\delta \hat{o}~(\hat{o}={n}, {m}, \hat{q}, \hat{p})$. Substituting these into Eq.~\ref{3E} and neglecting second-order fluctuation terms, we obtain the following steady-state values
\begin{equation}
	\begin{aligned}
		m_s &= \frac{\varepsilon_l \,(i \Delta_n + \kappa_n)}
		{\mathcal{J}^2 + (i \Delta_n + \kappa_n)\,(i \tilde{\Delta}_m + i \Delta_B + \kappa_m)}, ~
		p_s = 0, ~ 
		q_s  = - \frac{g\, |m_s|^2}{\omega_b}, \quad \\
		n_s &= - \frac{\mathcal{J}\, m_s}{\Delta_n^2 - 4 \chi^2}\,
		(\Delta_n + i\, 2 \chi e^{i \beta}),
	\end{aligned}
\end{equation} 
 where $\tilde{\Delta}_m=\Delta_m+g q_s$. The linearized QLEs describing the quadrature fluctuations $\left(\delta \hat{X}_n, \delta \hat{Y}_n, \delta \hat{X}_m, \delta \hat{Y}_m, \delta \hat{q}, \delta \hat{p}\right)$, with $\quad \delta \hat{X}_n=\left(\delta {n}+\delta {n}^{\dagger}\right) / \sqrt{2}, \delta \hat{Y}_n=i\left(\delta {n}^{\dagger}-\delta {n}\right) / \sqrt{2}, \delta \hat{X}_m=$ $\left(\delta {m}+\delta {m}^{\dagger}\right) / \sqrt{2}$, and $\delta \hat{Y}_m=i\left(\delta {m}^{\dagger}-\delta {m}\right) / \sqrt{2}$, can be written in a compact form as $\dot{u}(t)=\mathcal{A} u(t)+f(t)$
 where $u(t) = \Big[  \delta \hat{X}_n(t),\ \delta \hat{Y}_n(t),\ \delta \hat{X}_m(t),  \delta \hat{Y}_m(t),\ \delta \hat{q}(t),\ \delta \hat{p}(t) \Big]^{\mathrm{T}}$
 and 
$f(t) = \Big( \sqrt{2 \kappa_n} \, \delta \hat{X}_a^{\text{in}}(t),\ 
	\sqrt{2 \kappa_n} \, \delta \hat{Y}_n^{\text{in}}(t),\ 
	\sqrt{2 \kappa_m} \, \delta \hat{X}_m^{\text{in}}(t), \\
	 \sqrt{2 \kappa_m} \, \delta \hat{Y}_m^{\text{in}}(t),\ 
	0,\ 
	\xi(t) \Big)^{\mathrm{T}}$
  with $\quad \delta \hat{X}_n^{\text {in }}=\left(n^{\text {in }}+n^{\text {in } \dagger}\right) / \sqrt{2}, \hat{Y}_n^{\text {in }}=i\left({n}^{\text {in } \dagger}-\right.$ $\left.{n}^{\text {in }}\right) / \sqrt{2}, \delta \hat{X}_m^{\text {in }}=\left(m^{\text {in }}+ m^{\text {in } \dagger}\right) / \sqrt{2}, \quad$ and $\quad \hat{Y}_m^{\text {in }}=i\left({m}^{\text {in } \dagger}-\right.$ $\left.{m}^{\text {in }}\right) / \sqrt{2}$ being the vectors of input noises. The coefficient
matrix $\mathcal{A}$ is given by
\begin{widetext}  
	\begin{eqnarray} 
		\mathcal{A}=\begin{pmatrix}
			-\kappa_n +2 \chi \cos \beta & \Delta_n+ 2 \chi \sin \beta & 0 & \mathcal{J} & 0 & 0 \\
			-\Delta_n + 2 \chi \sin \beta & -\kappa_n-2 \chi \cos \beta & -\mathcal{J} & 0 & 0 & 0 \\
			0 & \mathcal{J} & -\kappa_m & \tilde{\Delta}_m+\Delta_B & -{\mathcal{G} } & 0 \\
			-\mathcal{J} & 0 & -\tilde{\Delta}_m-\Delta_B & -\kappa_m & 0 & 0 \\
			0 & 0 & 0 & 0 & 0 & \omega_b \\
			0 & 0 & 0 & {\mathcal{G} } & -\omega_b & -\gamma_b \end{pmatrix},
\end{eqnarray}
\end{widetext}
where ${\mathcal{G}}=\sqrt{2} \mathrm{igm}_s$ is the effective magnomechanical coupling rate. It is worth noting that the steady-state value $\left\langle d_s\right\rangle$ can be written in the simpler form 
\begin{equation}
m_s = \frac{i \varepsilon_l \Delta_n}{\mathcal{J}^2 - \Delta_n\left(\tilde{\Delta}_m + \Delta_B\right)}.
\label{drive}
\end{equation}
In the regime where $|\tilde{\Delta}_m|, |\Delta_n| \gg \kappa_n, \kappa_b, \chi$, the steady-state amplitude $m_s$ becomes purely imaginary, leading to approximately real effective coupling strengths ${\mathcal{G}}$ \cite{20}.
The steady state of the system is a zero-mean Gaussian state, fully characterized by a $6 \times 6$ covariance matrix (CM) $\mathcal{V}$, whose elements are defined as

$$
\mathcal{V}_{ij} = \frac{1}{2} \left\langle u_i(\infty) u_j(\infty) + u_j(\infty) u_i(\infty) \right\rangle, \quad (i, j = 1, 2, \ldots, 6).
$$
 To obtain the steady-state covariance matrix, one solves the corresponding Lyapunov equation \cite{108} 
 \begin{equation}
\mathcal{A V}+\mathcal{V} \mathcal{A}^{\mathrm{T}}+\mathcal{D}=0,
 \end{equation}
  where $\mathcal{D}$ is the diffusion matrix defined by $\left\langle v_i(t) v_j\left(t^{\prime}\right)+v_j\left(t^{\prime}\right) v_i(t)\right\rangle / 2=\mathcal{D}_{i j} \delta\left(t-t^{\prime}\right)$ and can be written as $\mathcal{D}=\mathcal{D}_n \oplus \mathcal{D}_m \oplus \mathcal{D}_b$, with $\mathcal{D}_n=$ $\operatorname{diag}\left[\kappa_n, \kappa_n\right], \mathcal{D}_m=\operatorname{diag}\left[\kappa_m, \kappa_m \right]$, and $\mathcal{D}_b=\operatorname{diag}\left[0, \gamma_b \left(2 N_b +\right.\right.$ 1)]. 
\section{entanglement and nonreciprocal entanglement  measures }
\label{p4}
To analyze the bipartite entanglement between subsystems $(i, j = n, b, m; \, i \neq j)$, we employ the logarithmic negativity $\mathcal{E}_{ij}$ \cite{109,110}, defined as follows:

\begin{equation}
	\mathcal{E}_{{ij}} \equiv \max \left[0,-\ln 2 \tilde{v}_{i \mid j}\right],
\end{equation}
with  $\tilde{v}_{i \mid j}=\min \operatorname{eig}\left|\left[\oplus_{s=1}^2\left(-\sigma_y\right)\right] \mathcal{P}_{1 \mid 2} \mathcal{V}_4 \mathcal{P}_{1 \mid 2}\right| \quad\left(\sigma_y\right.$ is the $y$ Pauli matrix) is the minimum symplectic eigenvalue of the CM $\mathcal{V}_4$ $(4 \times 4$ $\mathrm{CM}$ of two subsystems $)$ and $\mathcal{P}_{1 \mid 2}=$ $\operatorname{diag}(1,-1,1,1)$.

To quantitatively characterize the nonreciprocal entanglement resulting from the Barnett effect, we introduce the bidirectional contrast ratio $\hat{N}_{ij}$ (ranging from 0 to 1) as a measure of nonreciprocity in bipartite entanglement. It is defined as

\begin{equation}
	\hat{N}_{ij} = 
	\frac{\left| \mathcal{E}_{{ij}}(\Delta_B>0) - \mathcal{E}_{{ij}}(\Delta_B<0) \right|}
	{\mathcal{E}_{{ij}}(\Delta_B>0) + \mathcal{E}_{{ij}}(\Delta_B<0)}.
\end{equation}

Here, $\mathcal{E}_{{ij}}(\Delta_B>0)$ and $\mathcal{E}_{{ij}}(\Delta_B<0)$ represent the logarithmic negativities when the YIG is driven in a positive and negative direction, respectively. A value of $\hat{N}_{ij} = 1$ corresponds to perfect nonreciprocity, whereas $\hat{N}_{ij} = 0$ indicates reciprocal behavior.

\section{Results and Discussions}\label{sec3}

We base our calculations on the following experimentally accessible parameters, which are consistent with previous studies \cite{15,16,20}: $\omega_n / 2\pi = 10\ \mathrm{GHz}$, $\omega_b / 2\pi = 10\ \mathrm{MHz}$, $\kappa_n / 2\pi = \kappa_m / 2\pi = 1\ \mathrm{MHz}$, $\gamma_b / 2\pi = 100\ \mathrm{Hz}$, ${\mathcal{J}} / 2\pi = 3.2\ \mathrm{MHz}$, ${\mathcal{G}} / 2\pi = 4.8\ \mathrm{MHz}$, and a cryogenic temperature of $T = 10\ \mathrm{mK}$. The system features a YIG microsphere of radius $R = 250\ \mu\mathrm{m}$, containing approximately $N \simeq 3.5 \times 10^{16}$ spins, and driven by a magnetic field $B_0 \simeq 4 \times 10^{-5}\ \mathrm{T}$. The corresponding single-spin coupling rate is $g / 2\pi = 0.2\ \mathrm{Hz}$ \cite{13}, leading to a drive amplitude $\varepsilon_l \simeq 7.1 \times 10^{14}\ \mathrm{Hz}$, and an effective coupling strength ${\mathcal{G}} / 2\pi = 4.8\ \mathrm{MHz}$. In the following analysis, we explore how both bipartite entanglement and its nonreciprocity depend on various key system parameters.\\
		\begin{figure}[htbp]
			\centering  
			\includegraphics[width=0.47\textwidth]{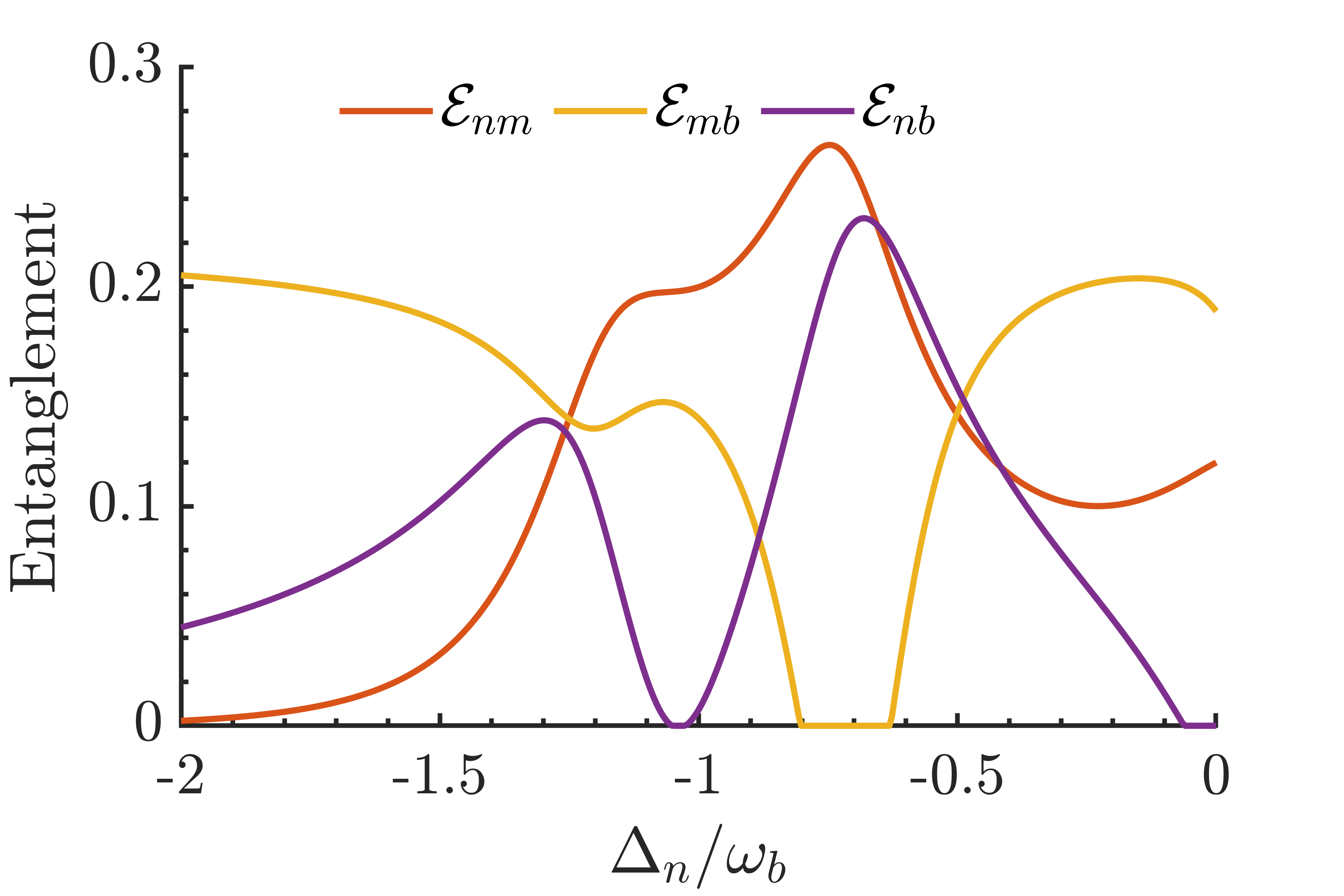} 
			\put(-60,152){\large(a)}  
			\hfil
			\includegraphics[width=0.47\textwidth]{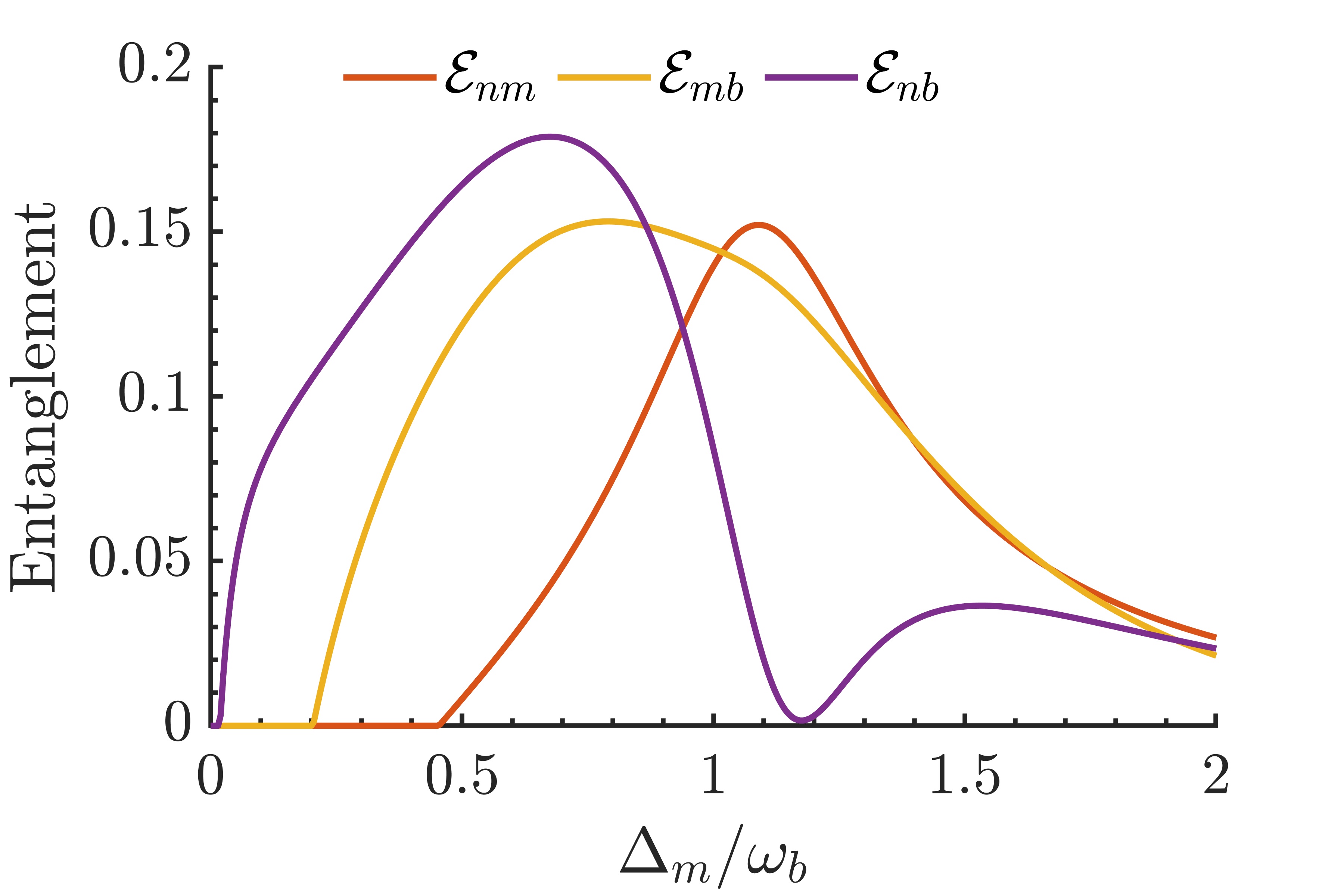} 
			\put(-60,152){\large(b)}  
			\caption{ Variation of logarithmic negativity $\mathcal{E}_{nm}$ between the microwave-magnon modes (the orange solid), $\mathcal{E}_{mb}$ between the magnon-phonon modes (the yellow solid), and $\mathcal{E}_{nb}$ between the microwave-phonon modes (the violet solid) as a function of the cavity mode detunings $\Delta_n$ (a) and the magnon mode detunings $\Delta_m$ (b), with $\chi=0.6 \kappa_n$ and $\beta=\pi$. The remaining parameters are detailed in the text.}
			\label{F1}
		\end{figure}
First, in Figure \ref{F1}, we present the entanglement variation between the microwave-magnon modes $\mathcal{E}_{nm}$, the magnon-phonon modes $\mathcal{E}_{mb}$, and the microwave-phonon modes $\mathcal{E}_{nb}$ as a function of the cavity mode detuning $\Delta_n$ (Figure \ref{F1}(a)) and the magnon mode detuning $\Delta_m$ (Figure \ref{F1}(b)). The entanglement can be controlled by adjusting $\Delta_n$ and $\Delta_m$. We observe that the three bipartitions do not attain their maximum value simultaneously. For instance, in Figure \ref{F1}(a), when the bipartitions $\mathcal{E}_{nm}$ and $\mathcal{E}_{nb}$ reach their maximum ($\mathcal{E}_{nm}\approx 0.26$ and $\mathcal{E}_{nb} \approx 0.22$), the entanglement $\mathcal{E}_{mb}$ vanishes. The entanglement in the system originates from magnetomechanical interactions mediated by the magnetostrictive interaction. The entanglement gradually transfers from the magnon-phonon subsystem to the magnon-photon and photon-phonon subsystems \cite{20}. For $0 < \Delta_m/\omega_b < 1$ (Figure \ref{F1}(b)), the three bipartitions increase as $\Delta_m$ grows. However, when $\Delta_m/\omega_b$ is far from 1, the entanglement of the three bipartitions diminishes.
 
\begin{figure}[htbp]
	\centering  	 
	\includegraphics[width=0.43\textwidth]{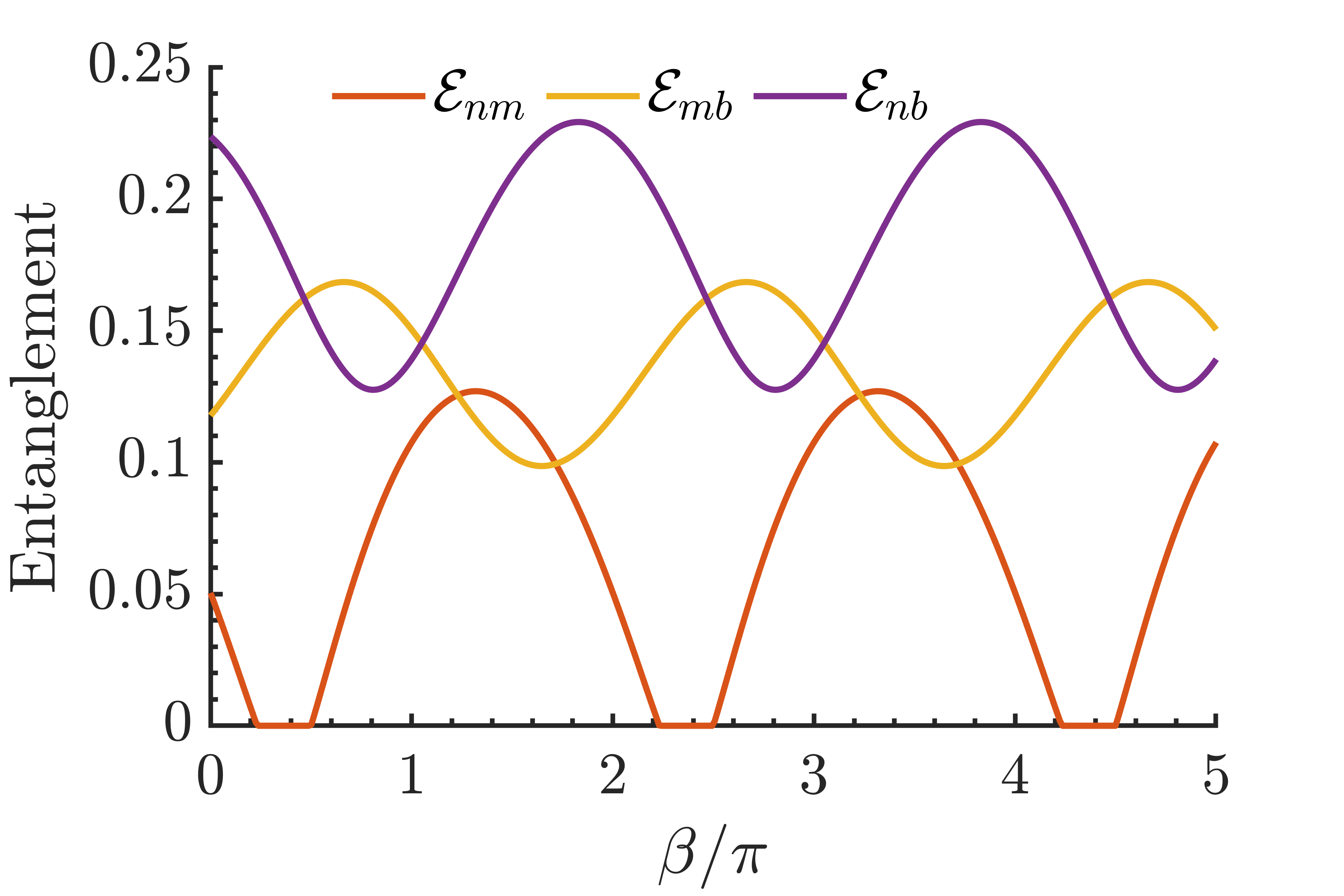}  
    	\put(-50,130){\large(a)}  
		\hfil
	\includegraphics[width=0.43\textwidth]{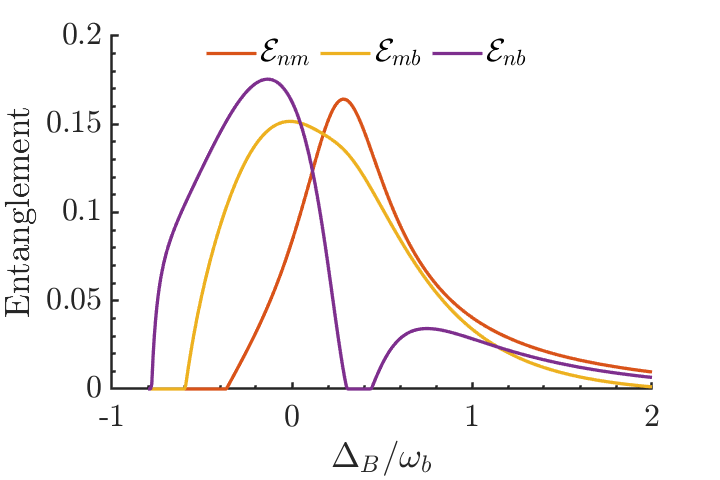}
	\put(-50,130){\large(b)} 
			\hfil
	\includegraphics[width=0.43\textwidth]{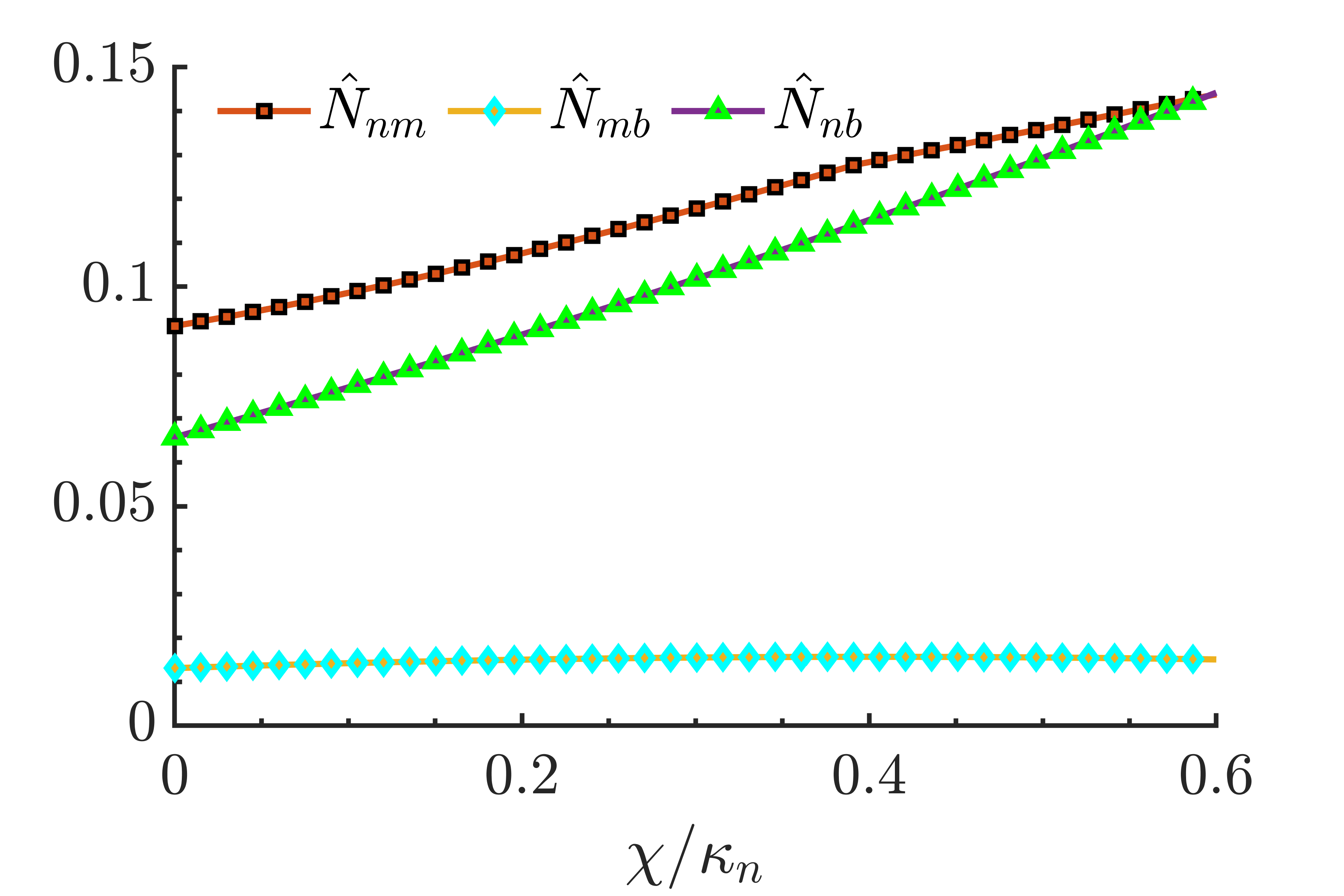}
		\put(-50,130){\large(c)} 
		\hfil
		\includegraphics[width=0.47\textwidth]{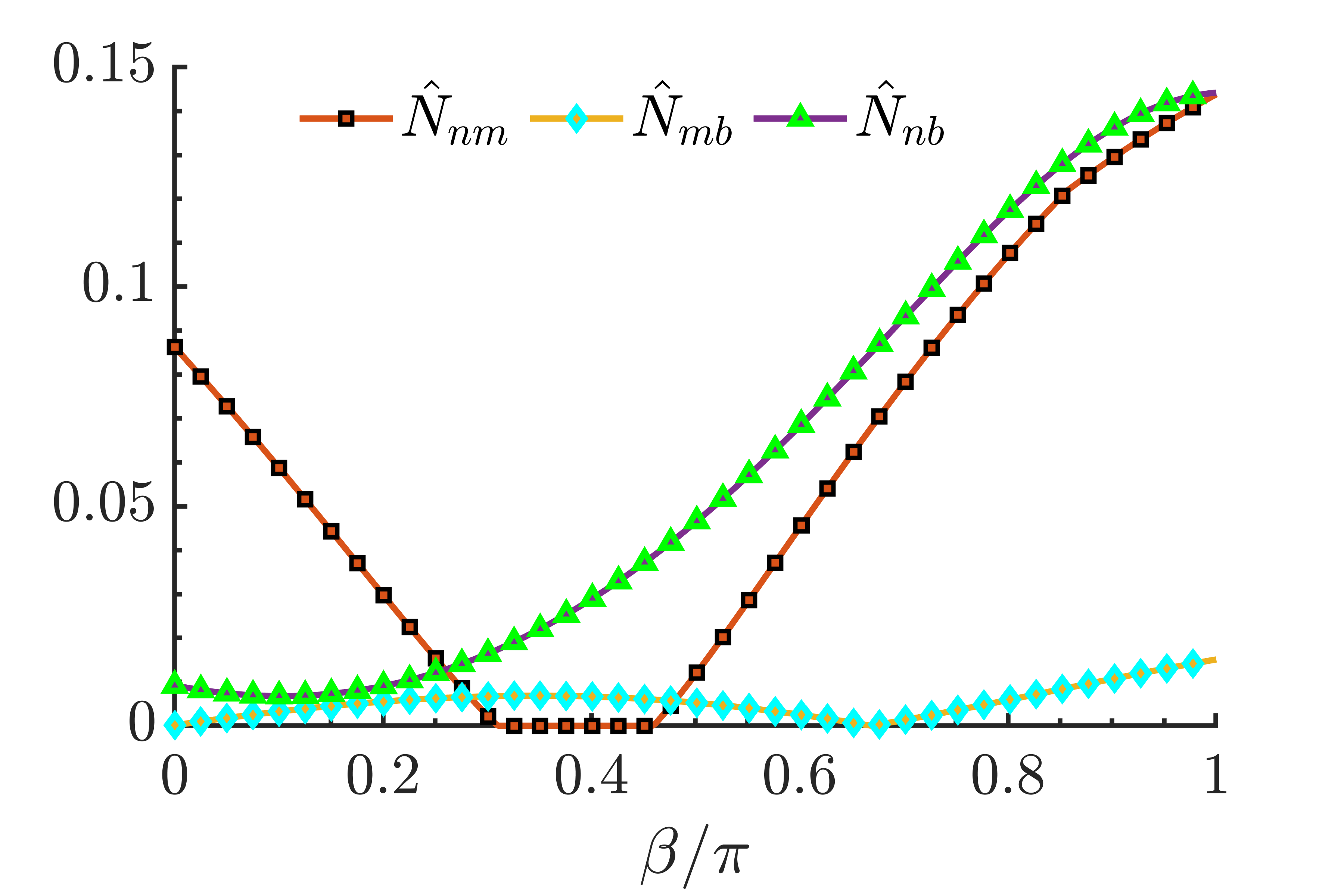}
	\put(-59,130){\large(d)} 		 
\caption{Plot of the logarithmic negativity $\mathcal{E}_{nm}$ between the microwave-magnon modes (the orange solid), $\mathcal{E}_{mb}$ between the magnon-phonon modes (the yellow solid), and $\mathcal{E}_{nb}$ between the microwave-phonon modes as a function of the phase shift of OPA (a), and the angular frequencies shift $\Delta_B$ (b). The nonreciprocity of entanglement $\hat{N}_{nm}$ between the microwave-magnon modes, $\hat{N}_{mb}$ between the magnon-phonon modes, and $\hat{N}_{nb}$ between the microwave-phonon modes as a function of the phase shift of OPA (a), the angular frequencies shift $\Delta_B$ (b), between the three bipartitions as a function of the nonlinear gain $\chi$ (c), and as a function of the phase shift $\beta$ (d). Where $\Delta_n = -1.3 \omega_b$. The other parameters are the same in Figure  \ref{F1}. }
\label{Fz}
\end{figure}
Figure \ref{Fz} (a) illustrates the three bipartitions of entanglement $\mathcal{E}_{nm}$, $\mathcal{E}_{mb}$, and $\mathcal{E}_{nb}$ as a function of the phase shift $\beta$ of the OPA. It is observed that all entanglements demonstrate a significant enhancement for a finely selective range of $\beta$. The magnon-phonon entanglement and the microwave-phonon entanglement appear to exhibit a periodic dependence on the phase $\beta$. For $\mathcal{E}_{nb}$, the entanglement is maximized when $\beta = (2n + 9/5)\pi$ (where $n \in \mathbb{Z}$) and is minimized at $\beta = (2n + 3/5)\pi$. Figure \ref{Fz}(b) illustrates the three bipartitions of entanglement $\mathcal{E}_{nm}$, $\mathcal{E}_{mb}$, and $\mathcal{E}_{nb}$ as a function of the normalized Barnett shift $\Delta_B$. It can be found that the maximal value of entanglement ($\mathcal{E}_{nm} \approx 0.16$ and $\mathcal{E}_{nb} \approx 0.18$) created for the two bipartitions (i.e., microwave-magnon and microwave-phonon) when $\Delta_B \neq 0$ is more important in comparison with the case where $\Delta_B = 0$ (where $\mathcal{E}_{nm} \approx 0.09$ and $\mathcal{E}_{nb} \approx 0.15$). We observe the simultaneous presence of an important amount of entanglement among the three bipartitions for $\Delta_n = -1.3 \omega_b$, which is not the same as that observed in Figure \ref{F1} (a) around this point (i.e., $\Delta_n = -1.3 \omega_b$). Figure \ref{Fz} (c) illustrates the variation in the nonreciprocity of entanglement between the microwave-magnon mode $\hat{N}_{nm}$, the magnon-phonon mode $\hat{N}_{mb}$, and the photon-phonon mode $\hat{N}_{nm}$ as a function of the normalized gain $\chi$.
The figure shows the dependence between the nonlinear gain (which is the coupling between the OPA and the cavity mode) and the nonreciprocity. The nonreciprocity increases as the gain $\chi$ increases, except for $\hat{N}_{nm}$, which exhibits a small enhancement as the gain increases. The nonreciprocity can also be manipulated by adjusting the OPA phase shift, as depicted in Figure \ref{Fz}(d). The nonreciprocity of all three bipartitions reaches its maximum for $\beta = \pi $. The phase $\beta $ can serve as a factor that improves the amount of nonreciprocity, so by controlling  $\beta$, the behavior of entanglement between the three bipartitions becomes more asymmetric.

\subsection{Magnomechanical coupling effect}
		\begin{figure}[htbp]
	\centering  
	\includegraphics[width=0.47\textwidth]{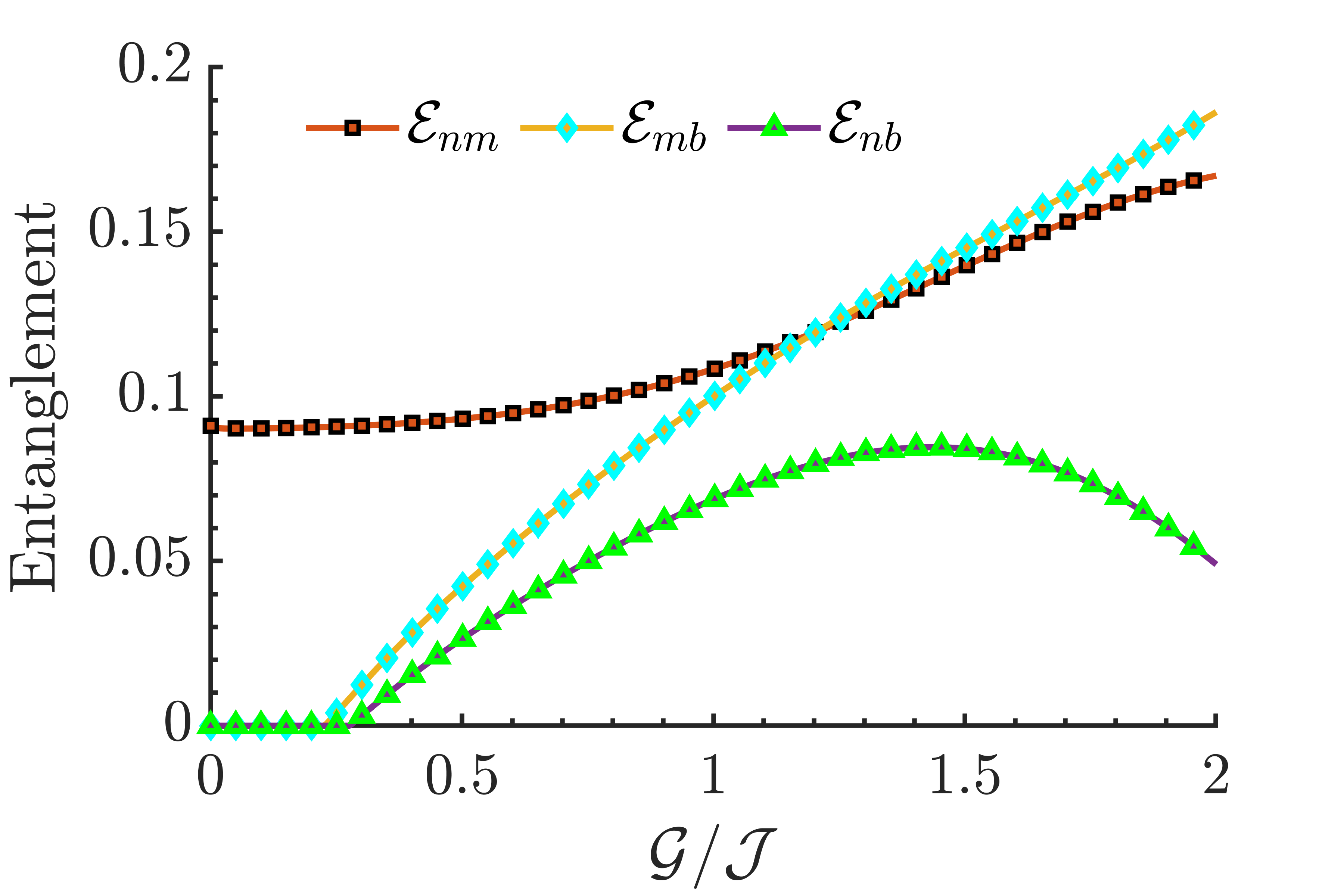}
	\put(-50,130){\large(a)} 
	\hfil 
	\includegraphics[width=0.47\textwidth]{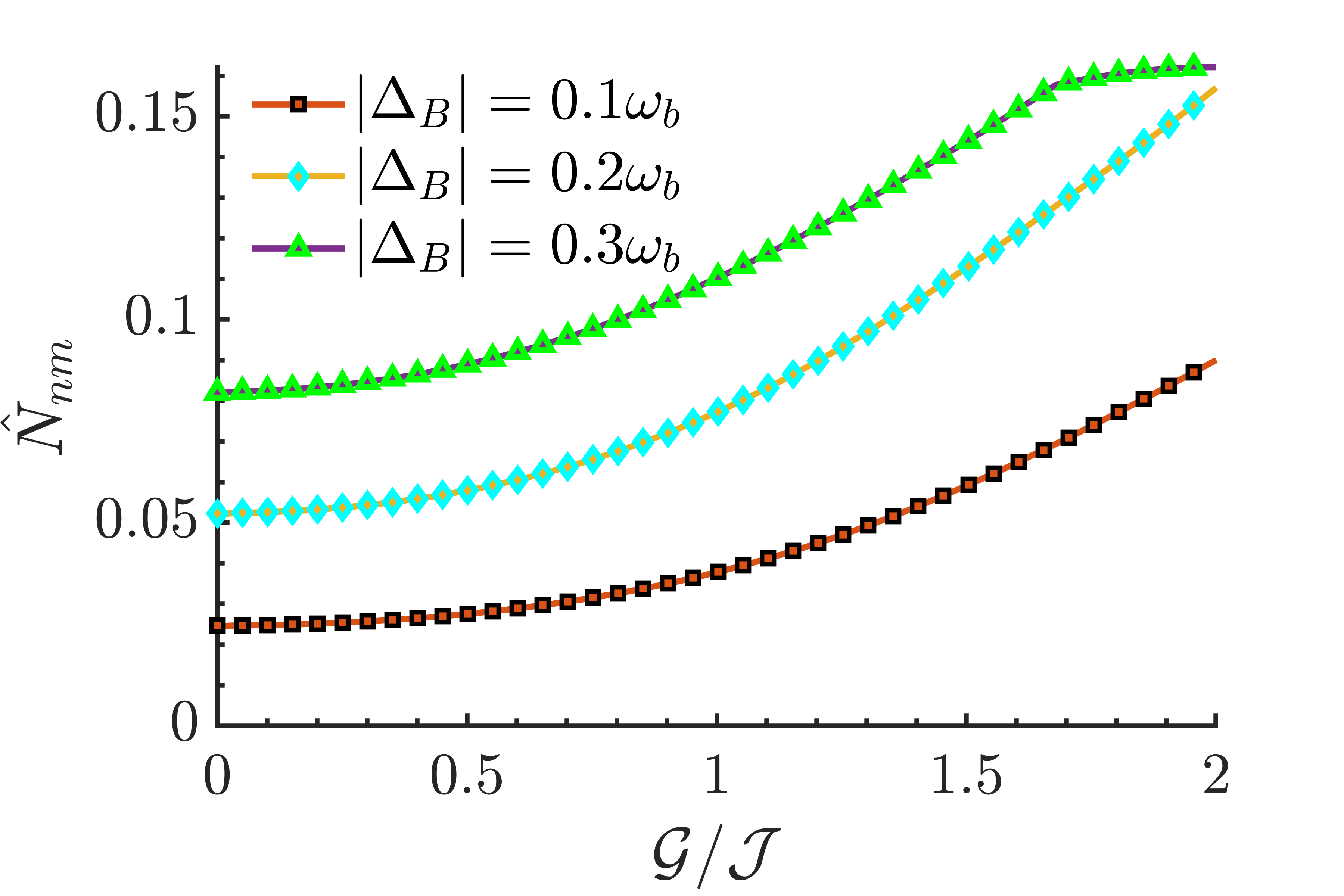}
	\put(-50,130){\large(b)} \\
	\hfil
	\includegraphics[width=0.51\textwidth]{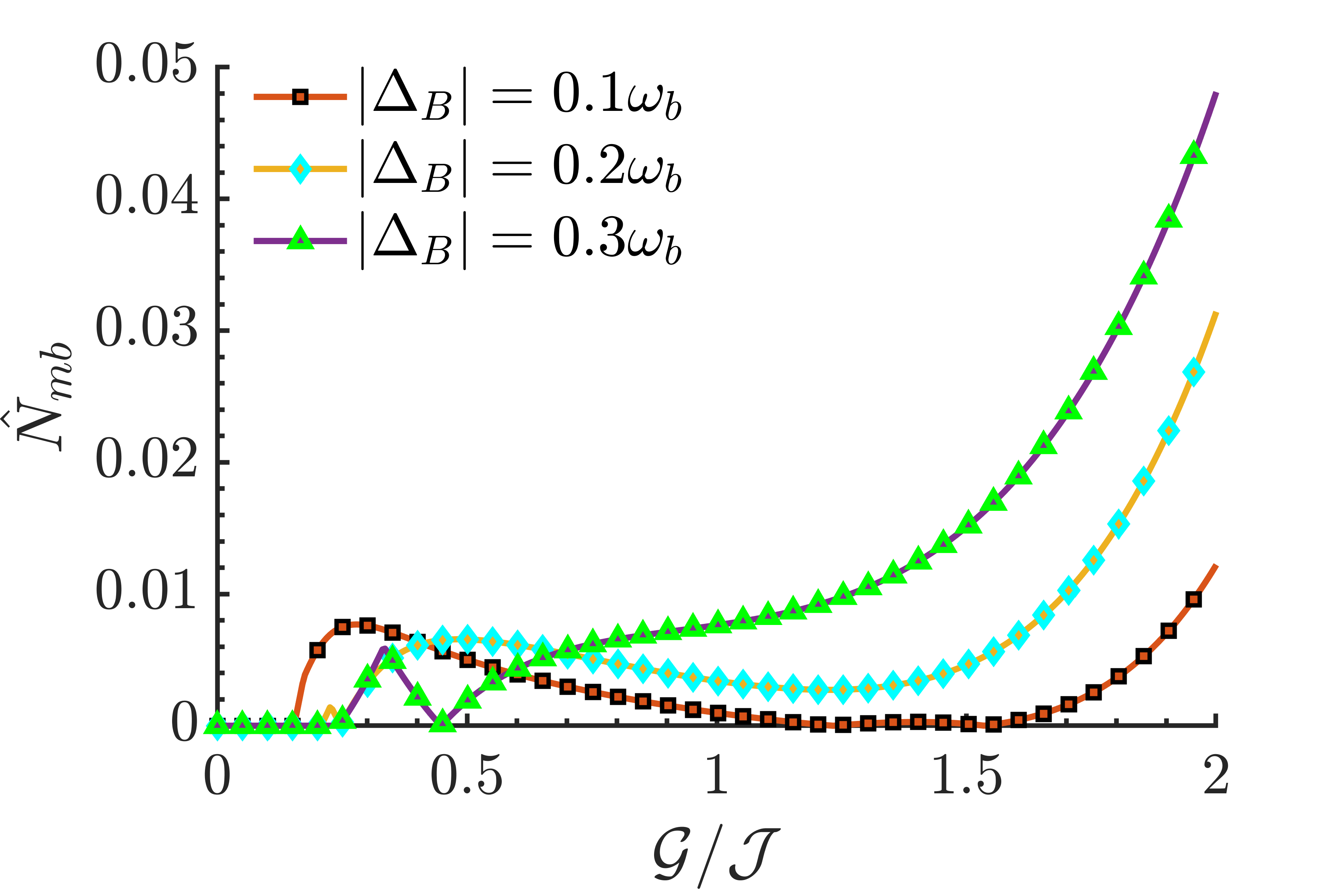}
	\put(-50,130){\large(c)} 
	\hfil
	\includegraphics[width=0.47\textwidth]{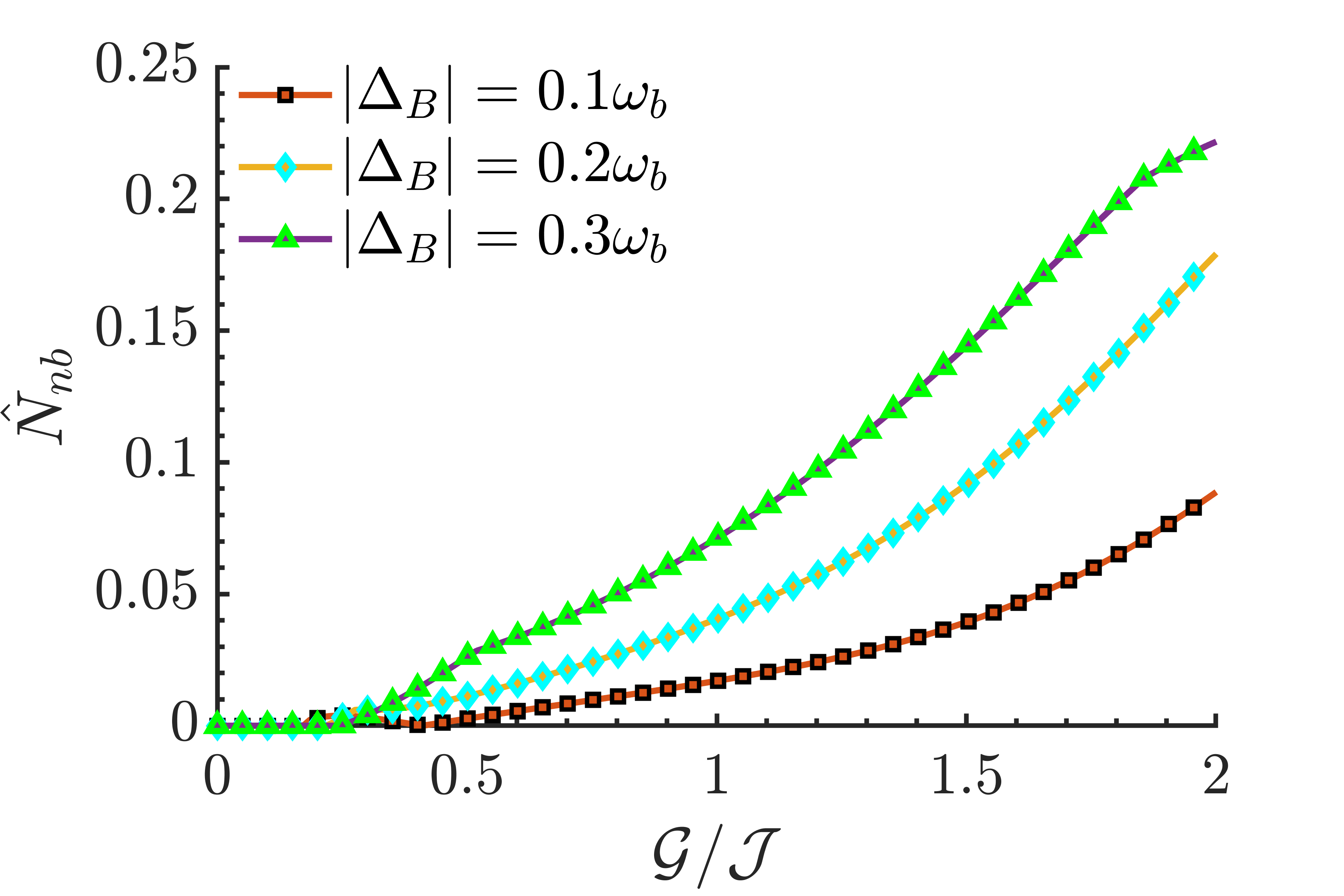}
	\put(-50,130){\large(d)} 
	\caption{(a) Plot of the logarithmic negativity $\mathcal{E}_{nm}$, $\mathcal{E}_{mb}$, and $\mathcal{E}_{nb}$ as a function of the normalized magnon-phonon coupling $\mathcal{J}$.  (b) The variations of the nonreciprocity of entanglement between the cavity mode and the magnon mode $\hat{N}_{nm}$, (c) between the magnon mode and the phonon mode $\hat{N}_{mb}$, and between the cavity mode  and the phonon mode $\hat{N}_{nb}$ (d) as a function of the normalized magnon-phonon coupling $\mathcal{J}$. With $\Delta_B=0.3 \omega_b$.}
	\label{F5}
\end{figure}
In the following, we aim to examine the nonreciprocity of entanglement through the bidirectional contrast ratio $\hat{N}$  and entanglement as a function of the magnomechanical coupling $\mathcal{G}$.
In Figure \ref{F5}, we plot the entanglement $\mathcal{E}_{mb}$, $\mathcal{E}_{nm}$, and $\mathcal{E}_{nb}$   (Figure \ref{F5}(a)),  and the nonreciprocity of entanglement $\hat{N}_{nm}$, $\hat{N}_{mb}$, and $\hat{N}_{nb}$,  between the three bipartitions as a function of the magnon-phonon coulpling $\mathcal{G}$.   
In fact, the magnomechanical coupling strength ${\mathcal{G}}$ can be fine-tuned by adjusting the drive strength $\varepsilon_l$ on the cavity [see Eqs.\ref{drive}] in our proposal. The range of regions in which nonreciprocal entanglement arises and is strongly dependent on the magnon-phonon coupling. Generally we observe that nonreciprocity of entanglement and the entanglement increases with the increase of the ratio ${\mathcal{G}}/\mathcal{J}$, after surpassing a certain threshold of ${\mathcal{G}}/\mathcal{J}$, where the magnonmachanicals coupling ${\mathcal{G}}$  is more dominant than the magnon-photon  ${\mathcal{J}}$. For a fixed value of the magnomechanical coupling, we observe that the nonreciprocity among all bipartitions is enhanced by increasing the angular frequency values  $\Delta_B$  from  $\Delta_B = 0.1 \omega_b$ to $\Delta_B = 0.3 \omega_b$ . 

\subsection{Nonreciprocal bipartite entanglement}

\begin{figure}[htbp]
	\centering  
	\includegraphics[width=0.42\textwidth]{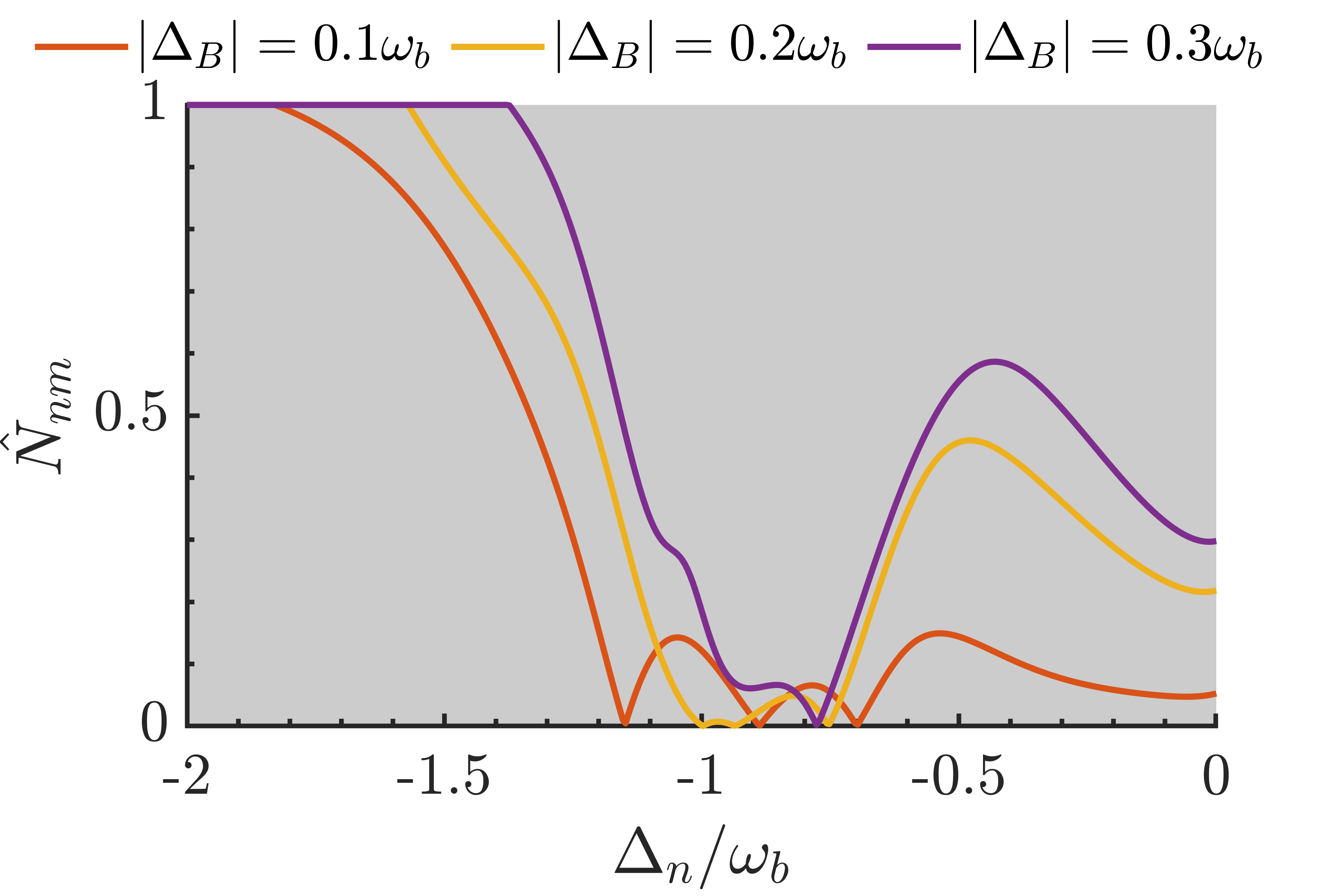}
	\put(-40,110){\large(a)} 
	\hfil
	\includegraphics[width=0.45\textwidth]{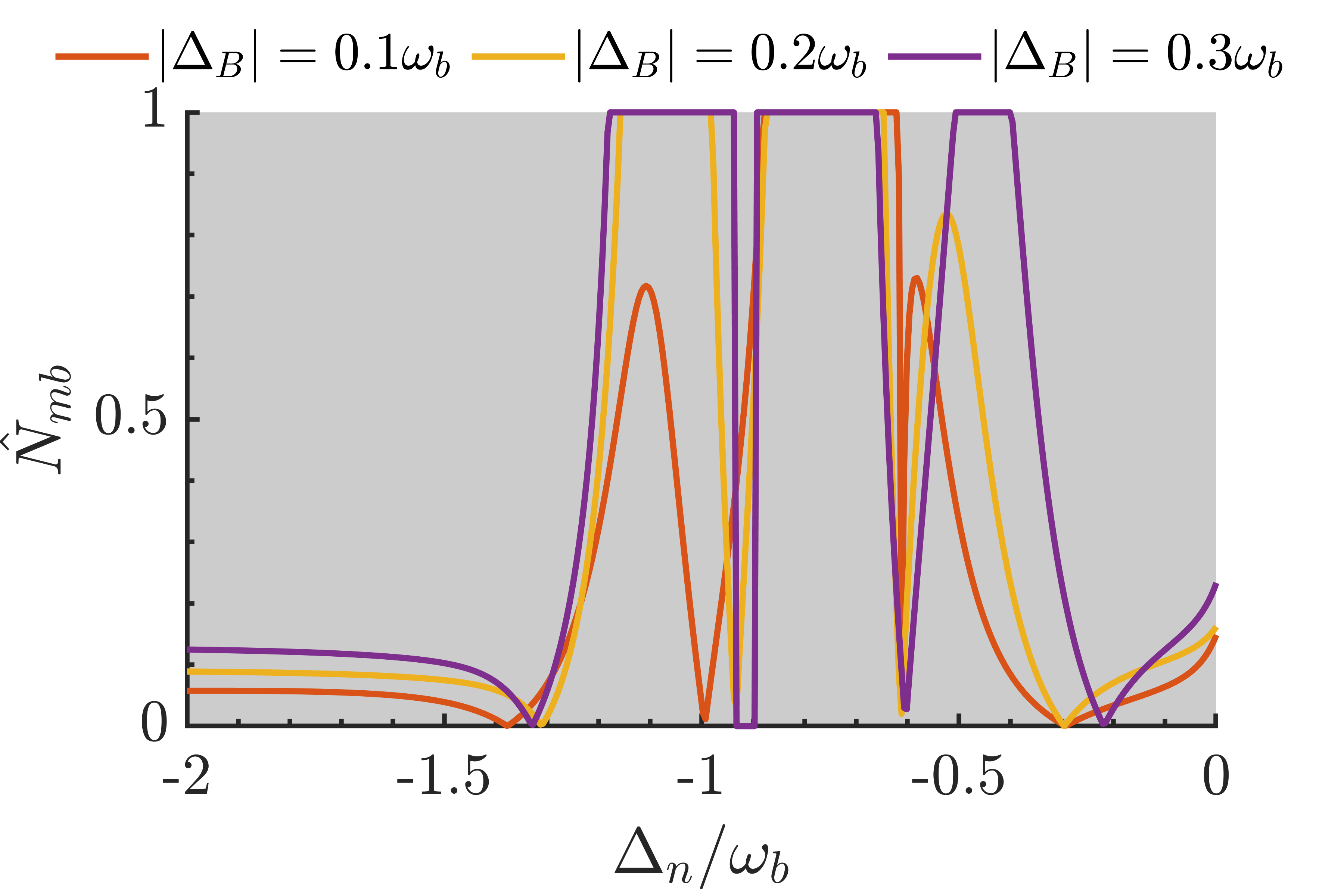}
	\put(-40,110){\large(b)}\\
	\includegraphics[width=0.45\textwidth]{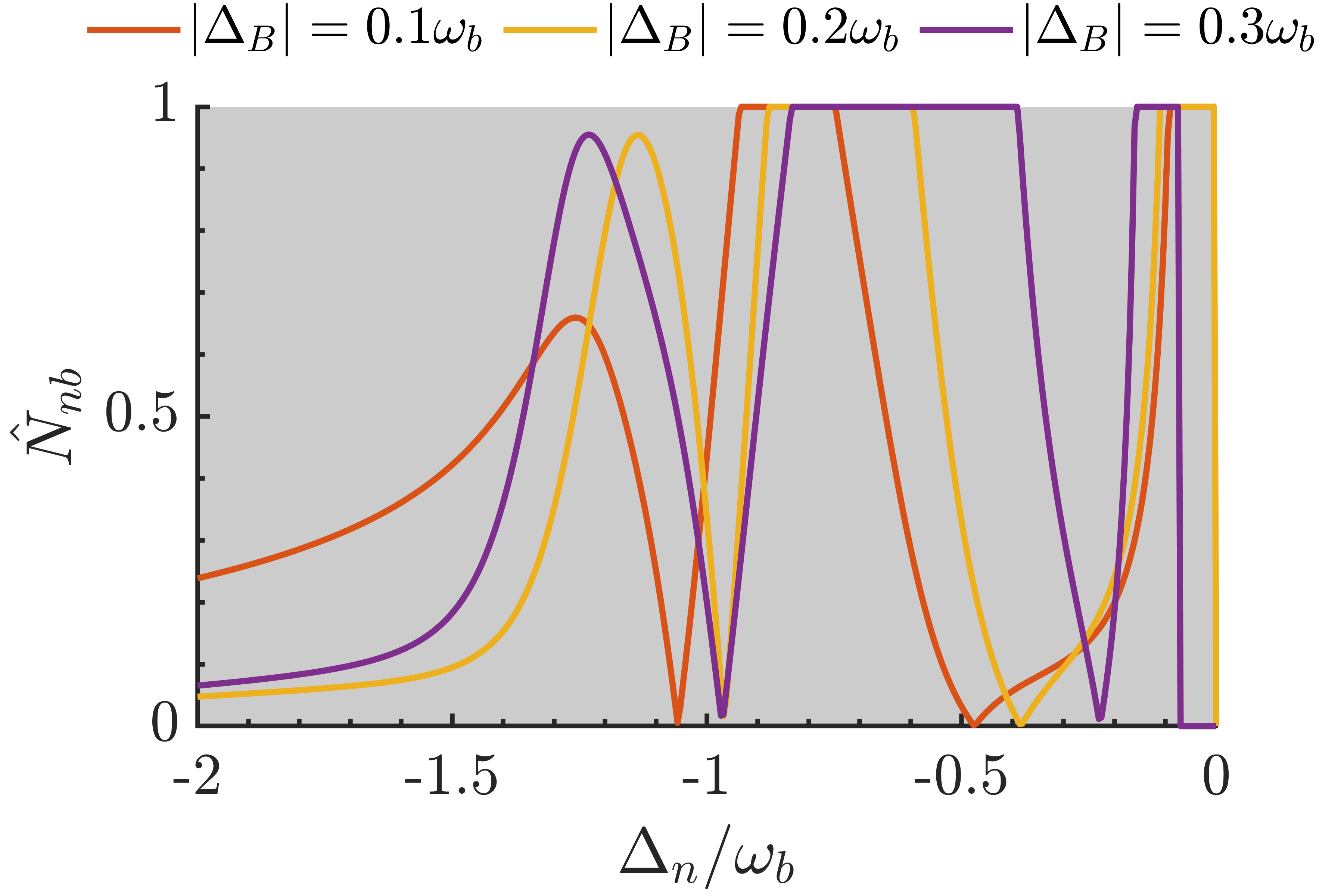}
	\put(-40,110){\large(c)}
	\caption{ Bidirectional contrast ratio $\hat{N}$ between the photon and the magnon modes $\hat{N}_{nm}$ (a), the magnon and the phonon modes $\hat{N}_{mb}$ (b), and between the photon and phonon modes $\hat{N}_{nb}$ (c), as a function of photon frequency detuning. The other parameters are the same as those in Figure \ref{F1}.}
\label{F50}
\end{figure}

\begin{figure}[htbp]
	\centering  
	\includegraphics[width=0.45\textwidth]{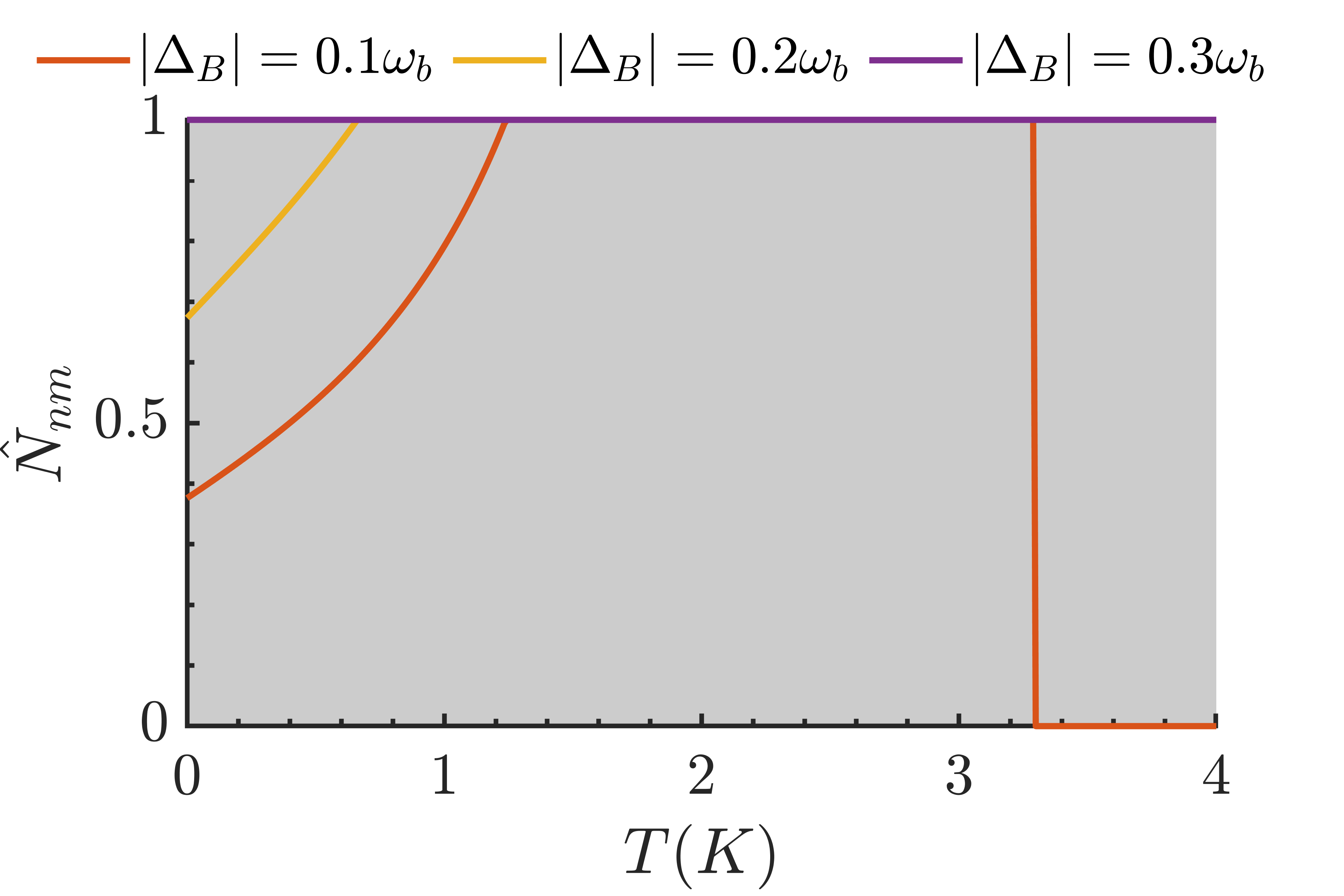}
       \put(-40,115){\large(a)} 
	\hfil
	\includegraphics[width=0.45\textwidth]{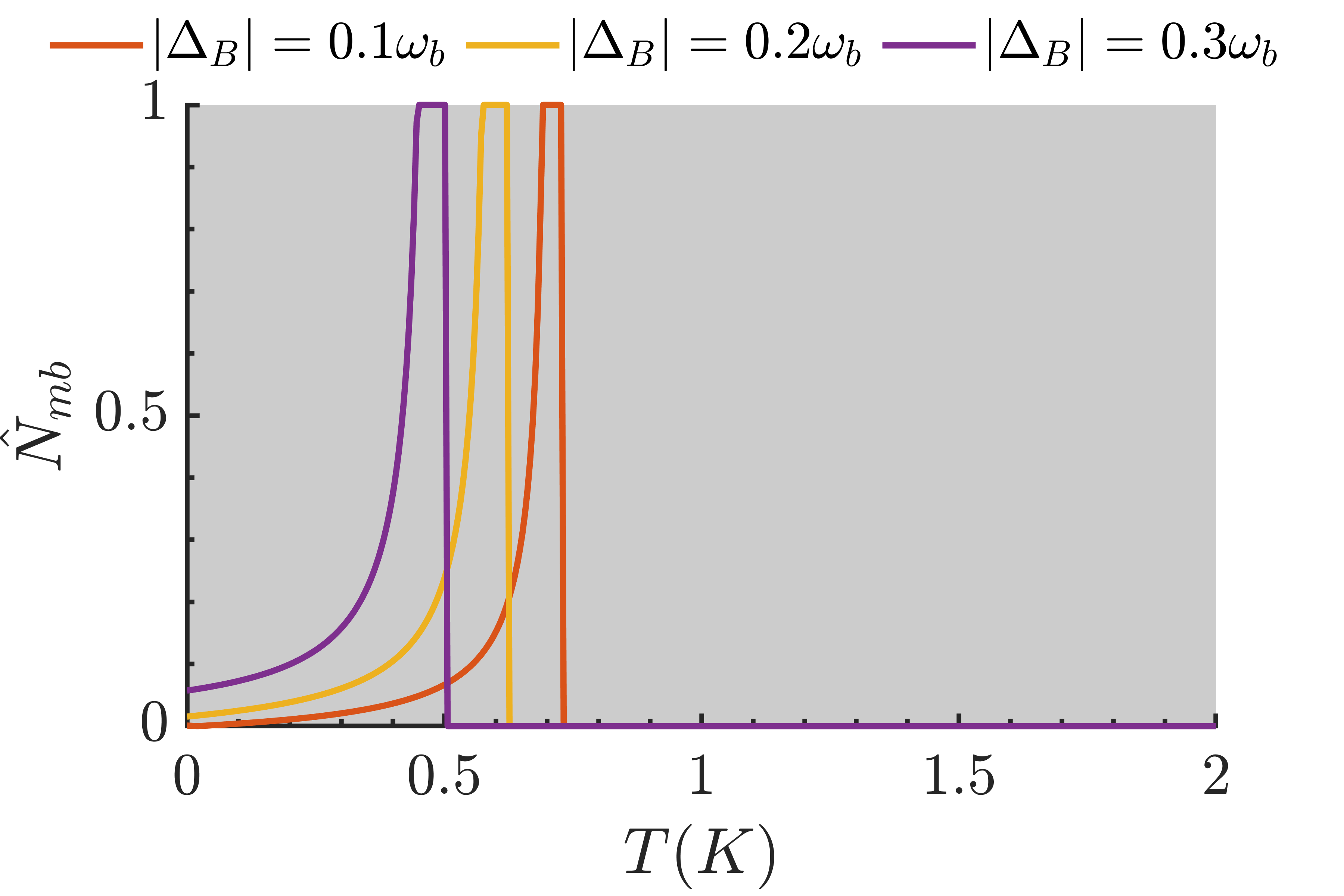}
       \put(-40,115){\large(b)}\\
	\includegraphics[width=0.45\textwidth]{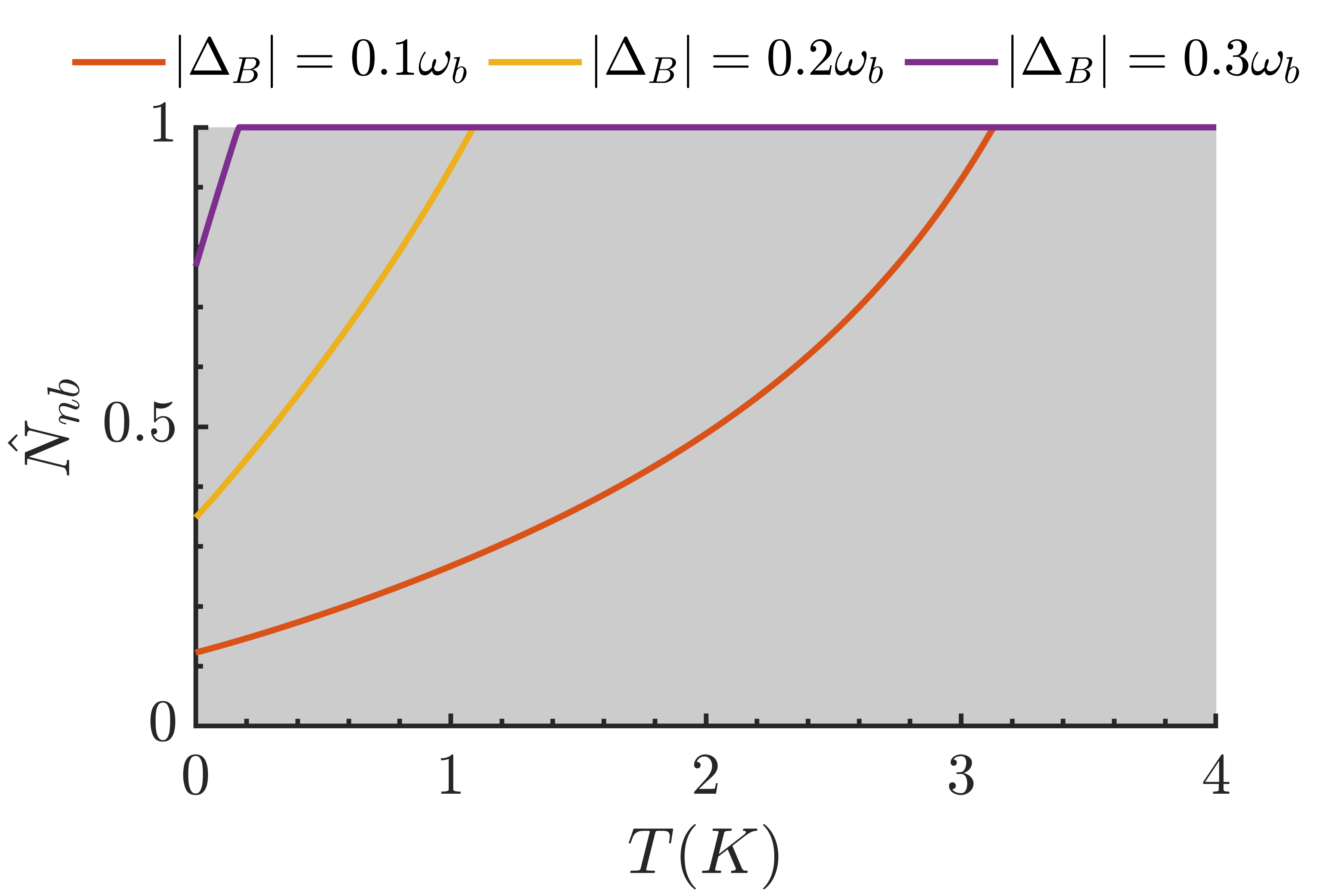}
	\put(-40,110){\large(c)}
	\caption{ Bidirectional contrast ratio $\hat{N}$ between the photon and the magnon modes $\hat{N}_{nm}$ (a), the magnon and the phonon modes  $\hat{N}_{mb}$ (b), and  between the photon and magnon modes $\hat{N}_{nb}$ (c), as a function of  bath temperature. The remaining parameters are consistent with those presented in Figure \ref{F1}.}
	\label{F503}
\end{figure}

A higher contrast ratio $\hat{N}$ indicates stronger entanglement nonreciprocity. Figures \ref{F50} and \ref{F503} illustrate the variation of the bidirectional contrast ratio $\hat{N}$  as a function of the normalized optical detuning  $\Delta_n/\omega_b$ ) for various values of  $\Delta_B$, and as a function of temperature, respectively. The Figures \ref{F50} clearly shows that the nonreciprocity of the bipartite entanglements can be actively controlled switched on or off, by tuning the photon detuning $\Delta_n/\omega_m$ , i.e., the bidirectional contrast ratio  can be tuned between $0$ and $1$ by varying $\Delta_n/\omega_m$, it is possible to achieve optimal nonreciprocal entanglement.  The optimal nonreciprocity is achieved in our scheme when $-2 < \Delta_{n} / \omega_d <-1.5$ for $\hat{N}_{nr}$ and when $-1.4 \leq \Delta_{n} / \omega_m \leq -0.9$ and $-0.9 \leq \Delta_{n} / \omega_m \leq -0.4$ for $\hat{N}_{mb}$, for $\hat{N}_{nm}$, we get a maximum nonreciprocity approximately when $-1 < \Delta_{n} / \omega_d <-0.5$. This shows that by cear adjusting the  frequency detuning $\Delta_{n}$, the entangled states are becoming more distinct and asymmetric in their behavior in different directions.
Nonreciprocity disappears only at a critical point of $\Delta_n$, where the entangled states become indistinguishable in both directions. The ability to switch between these two types of entanglement highlights the presence of a nonreciprocal mechanism in the entanglement generation process. This tunable control over entanglement is highly appealing for a wide range of quantum technologies. Furthermore, we observe an improvement over the results in \cite{bs}, highlighting the additional role of the OPA in enhancing nonreciprocity in entanglement, alongside its established effect on entanglement itself \cite{pr,pla}. Furthermore, the results indicate that operating at higher angular frequencies enhances the degree of entanglement nonreciprocity. 

The nonreciprocity of entanglement as a function of temperature variation (Figure \ref{F503}) reveals that, for all bipartitions, it gradually increases with rising temperature until reaches a maximum value. Our results demonstrate that significant nonreciprocity can be achieved even at relatively high temperatures, highlighting the potential of the proposed scheme for realizing robust and thermally resilient nonreciprocal quantum devices. Furthermore, we suggest that elevated temperatures can be advantageous for attaining large or optimal entanglement nonreciprocity, offering a promising avenue for engineering nonreciprocal behavior \cite{april2025}. As the Barnett shift $\Delta_B$ increases, the range where a ideal nonreciprocity appears becomes large.

\subsection{Barnett effect vs Temperature effect}
  
\begin{figure}[htbp]
	\centering  
	\includegraphics[width=0.44\textwidth]{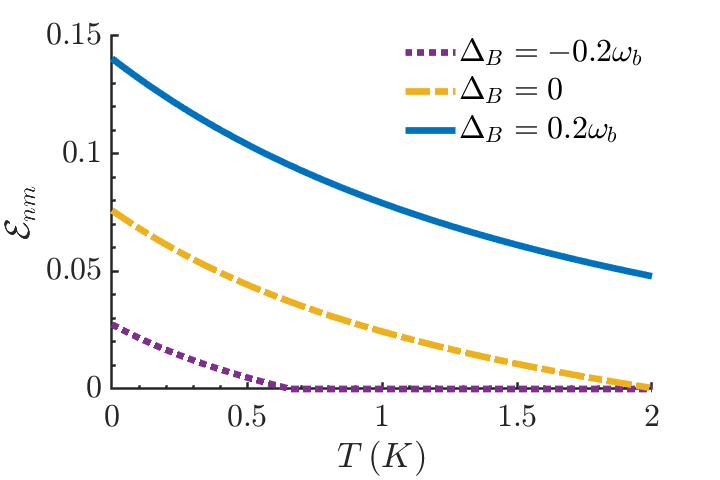}
	\put(-170,130){\large(a)} \\
	\includegraphics[width=0.44\textwidth]{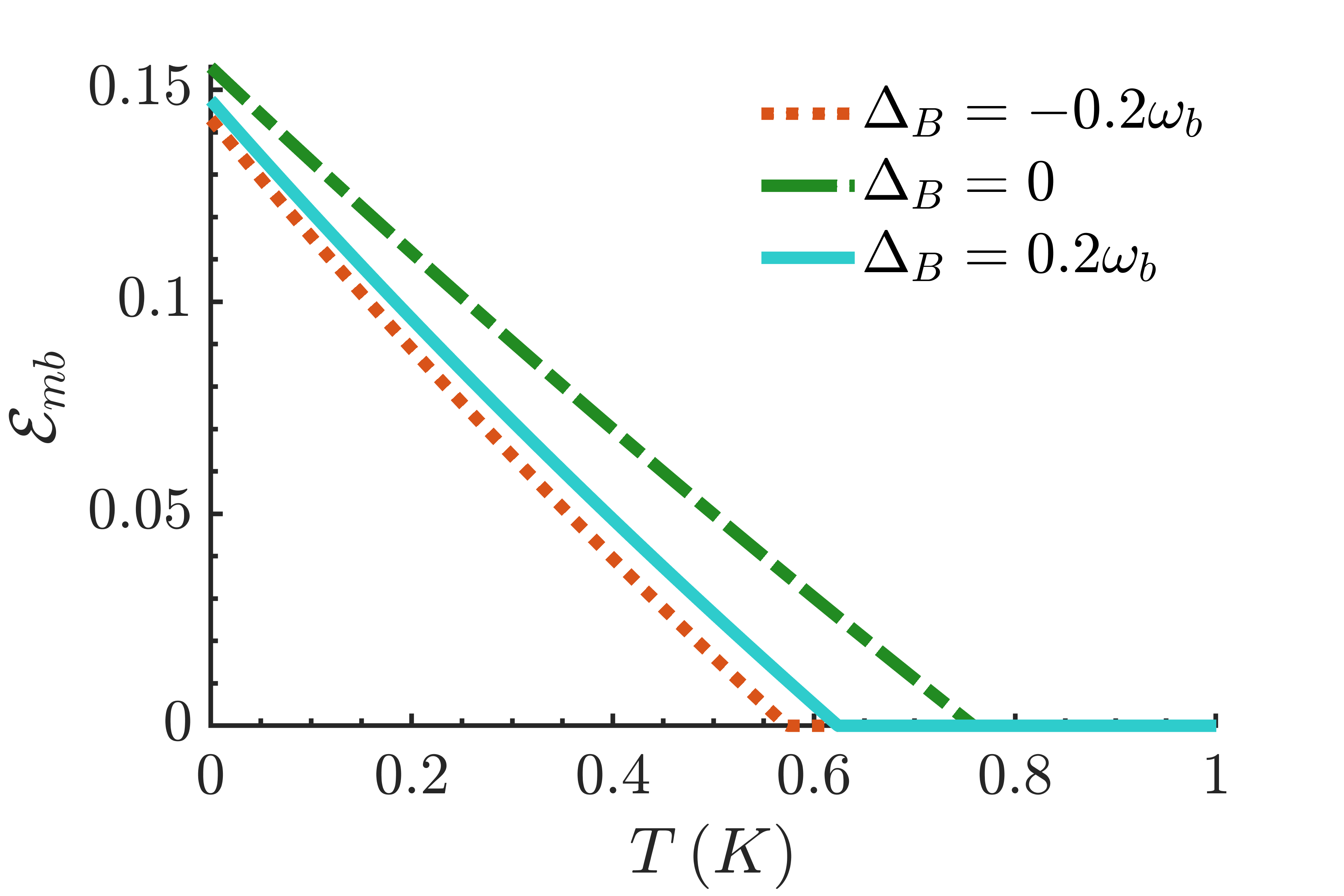}
	\put(-170,130){\large(b)}\\
	\includegraphics[width=0.44\textwidth]{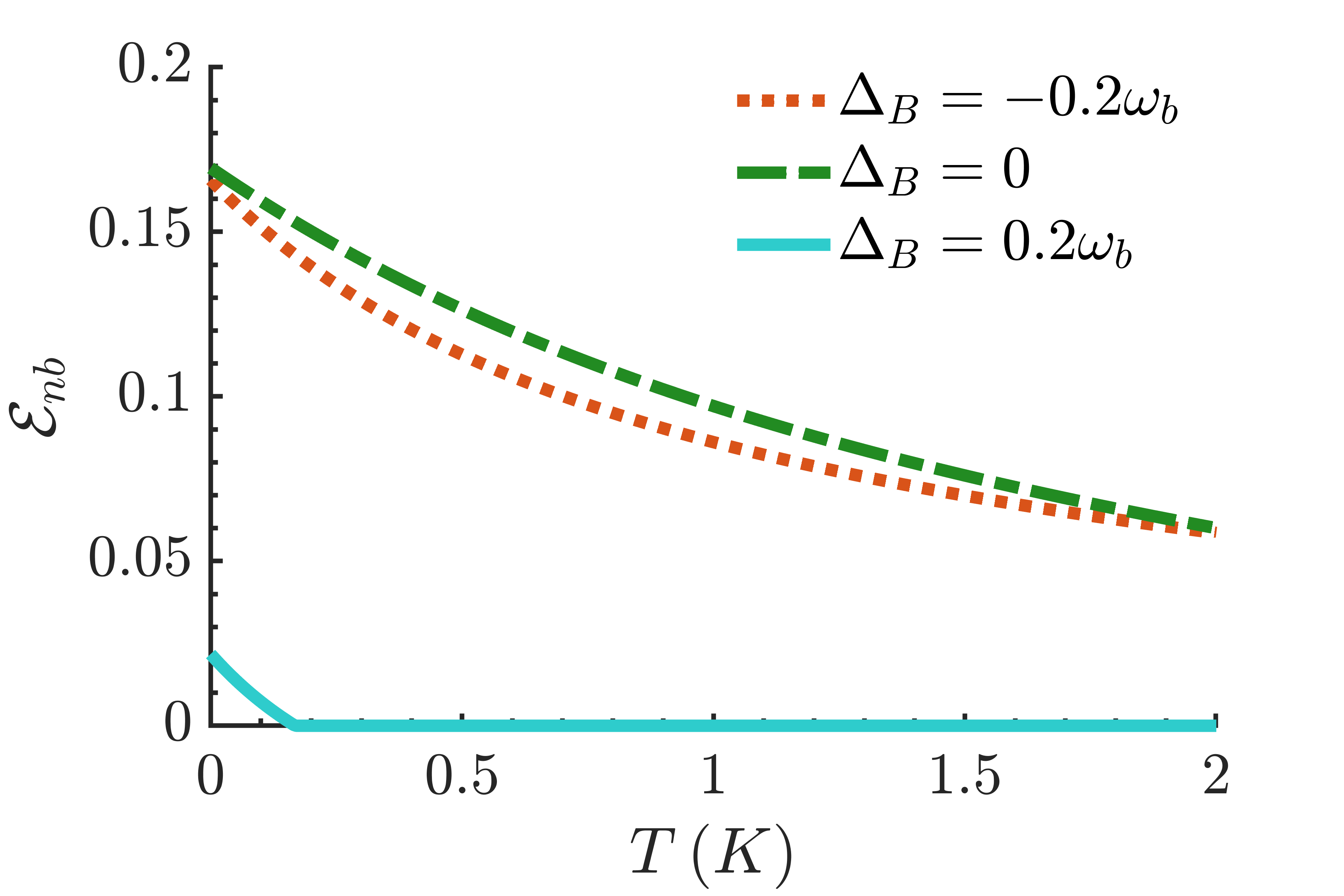}
    \put(-170,130){\large(c)}
	\caption{Plot of the logarithmic negativity $\mathcal{E}_{nm}$ (a), $\mathcal{E}_{mb}$ (b), and $\mathcal{E}_{nb}$ (c) as a function of temperature $T(K)$, for different value of the Barnett shift $\Delta_B$. The remaining parameters are consistent with those presented in Figure. \ref{F1}.}
	\label{F4T}
\end{figure}

Next, we investigate how the bath temperature influences the entanglement in our scheme. To this end, we plot the logarithmic negativity as a function of temperature $T$, for $ \Delta_B=0.2 \omega_b$, $ \Delta_B=0 \omega_b$, and $ \Delta_B=-0.2 \omega_b$. The entanglement decreases with the increase in temperature as a result of the decoherence induced by temperature. the Figure \ref{F4T} also shows the effect of the Barnett shift on making the nonreciprocal entanglement stronger in the face of temperature in comparison with \cite{bs}. The figure  illustrates that nonreciprocity exhibits a highly sensitive dependence on temperature. Previous studies have reported an upper temperature limit of $T = 1.4 \,\text{K}$ for achieving nonreciprocal entanglement \cite{bs}. This indicates that, in comparison to earlier works, nonreciprocal entanglement demonstrates remarkable robustness (persisting for a temperature above $T = 1.5 \,\text{K}$) against thermal  noise when the OPA is incorporated into CMM-based systems. The Figure \ref{F4T} also illustrates the previous conclusions about the asymmetry of entanglement when taking into account the Barnett effect. With the tuning of the direction of the magnetic field ($\pm$ z), we can achieve a significant improvement in bipartite entanglement.  Ideal nonreciprocity can be achieved across a broad range of temperature values, as previously demonstrated in Figure \ref{F503}. The microwave-phonon couplings exhibit very important nonreciprocity in comparison with the other bipartitions.
 \section{Nonreciprocity beyond entanglement}
 The covariance matrix  between two modes can always be put into a block form
 
 \begin{equation}
 	\sigma=\left(\begin{array}{cc}
 		\mathcal{B} & \mathcal{C} \\
 		\mathcal{C}^{\mathrm{T}} & \mathcal{D}
 	\end{array}\right),
 \end{equation}
 
 G. Adesso et al. proposed a method for calculating interferometric power of any two-mode Gaussian state \cite{gip}
 
 $$
 \mathcal{P}(\sigma)=\frac{U+\sqrt{U^2+R V}}{2 R},
 $$
 
 where
 
 $$
 \begin{aligned}
 	& U=\left(D_1+D_3\right)\left(1+D_2+D_3-D_\sigma\right)-D_\sigma^2 \\
 	& R=\left(D_\sigma-1\right)\left(1+D_1+D_2+2 D_3+D_\sigma\right) \\
 	& V=\left(D_1+D_\sigma\right)\left(D_1 D_2-D_\sigma\right)+D_3\left(2 D_1+D_3\right)\left(1+D_2\right)
 \end{aligned}
 $$
 
 with $D_\sigma=16 \operatorname{det}[\sigma], D_1= 4\operatorname{det}[\mathcal{B}], D_2= 4\operatorname{det}[\mathcal{D}]$, and $D_3=4\operatorname{det}[\mathcal{C}]$.

\begin{figure*}[htbp]
 	\centering  
 	\includegraphics[width=0.44\textwidth]{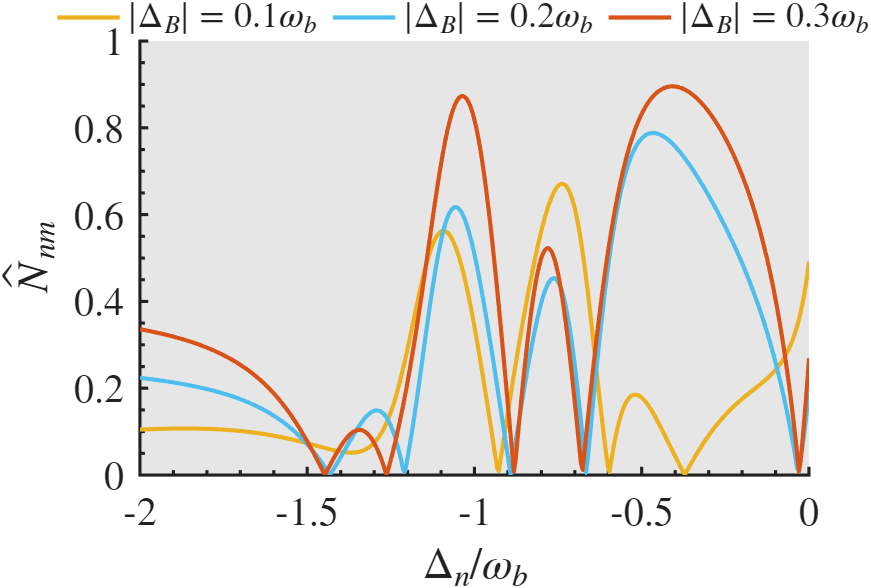}
 		\put(-40,130){\large(a)} 
 	\hfil
 	\includegraphics[width=0.44\textwidth]{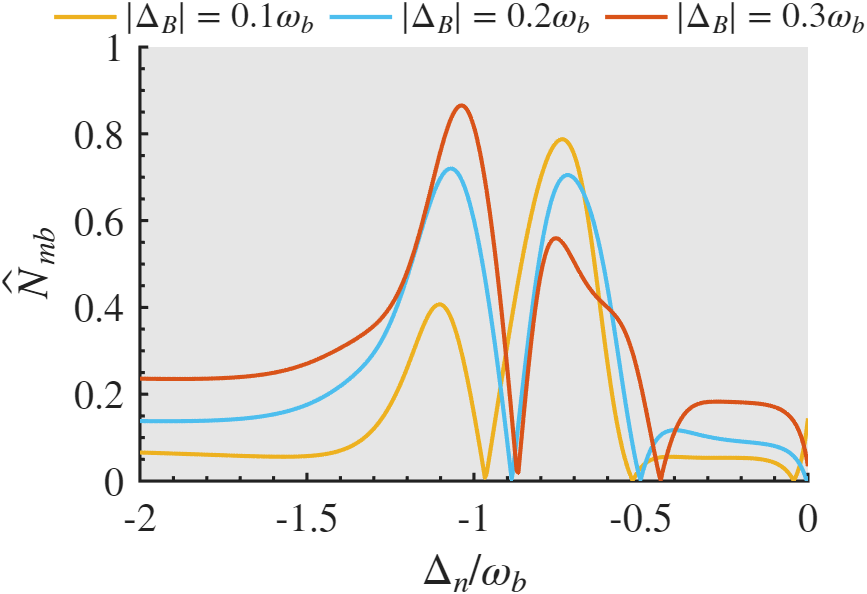}
     	\put(-40,130){\large(b)}
 	\hfil
 	\includegraphics[width=0.44\textwidth]{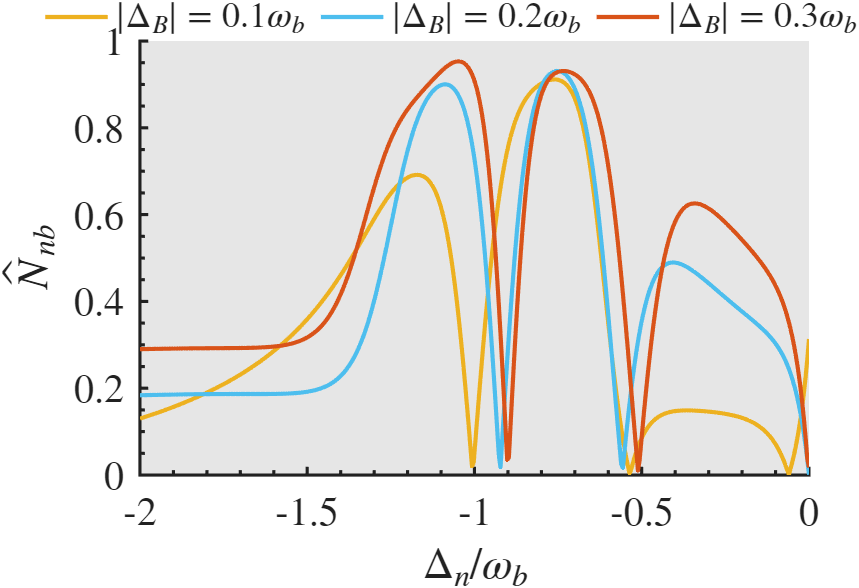}
 		\put(-40,130){\large(c)}
 	\caption{The nonriciprocity of the GIP in term of the contrast ration between the photon and the magnon modes $\hat{N}_{nm}$ (a), the magnon and the phonon modes  $\hat{N}_{mb}$ (b), and  between the photon and magnon modes $\hat{N}_{nb}$ (c), as a function of the detuning $\Delta_n$ of the cavity mode. The remaining parameters are consistent with those presented in Figure \ref{F1}. The remaining parameters are consistent with those presented in Figure \ref{F1}.}
 	\label{F9}
 \end{figure*}

 We studied the nonreciprocity of the GIP using the contrast ratio $\hat{N}$ defined as following
 \begin{equation}
 	\hat{N}_{ij} = 
 	\frac{\left| P_{{ij}}(\Delta_B>0) - P_{{ij}}(\Delta_B<0) \right|}
 	{P_{{ij}}(\Delta_B>0) + P_{{ij}}(\Delta_B<0)},
 \end{equation}
 where $\hat{N} = 1$ and $\hat{N} = 0$ correspond to perfect and no nonreciprocities for GIP, respectively. Hence, a larger value of $\hat{N}$ indicates a stronger degree of non-reciprocity in GIP, respectively. 
The Figure \ref{F9} illustrates the GIP as a function of the normalized cavity detuning $\Delta_n$. The GIP exhibits nonreciprocal behavior, which indicates that the GIP can respond differently when the direction of the applied magnetic field is changed. The nonreciprocity of GIP can be can be turned on or off by adjusting the cavity detuning  $\Delta_n$. By tuning the detuning $\Delta_n$, we can achieve the strongest nonreciprocity of GIP.

 \section{stability of the system }
\begin{figure}[htbp]
	\centering  
	\includegraphics[width=0.48\textwidth]{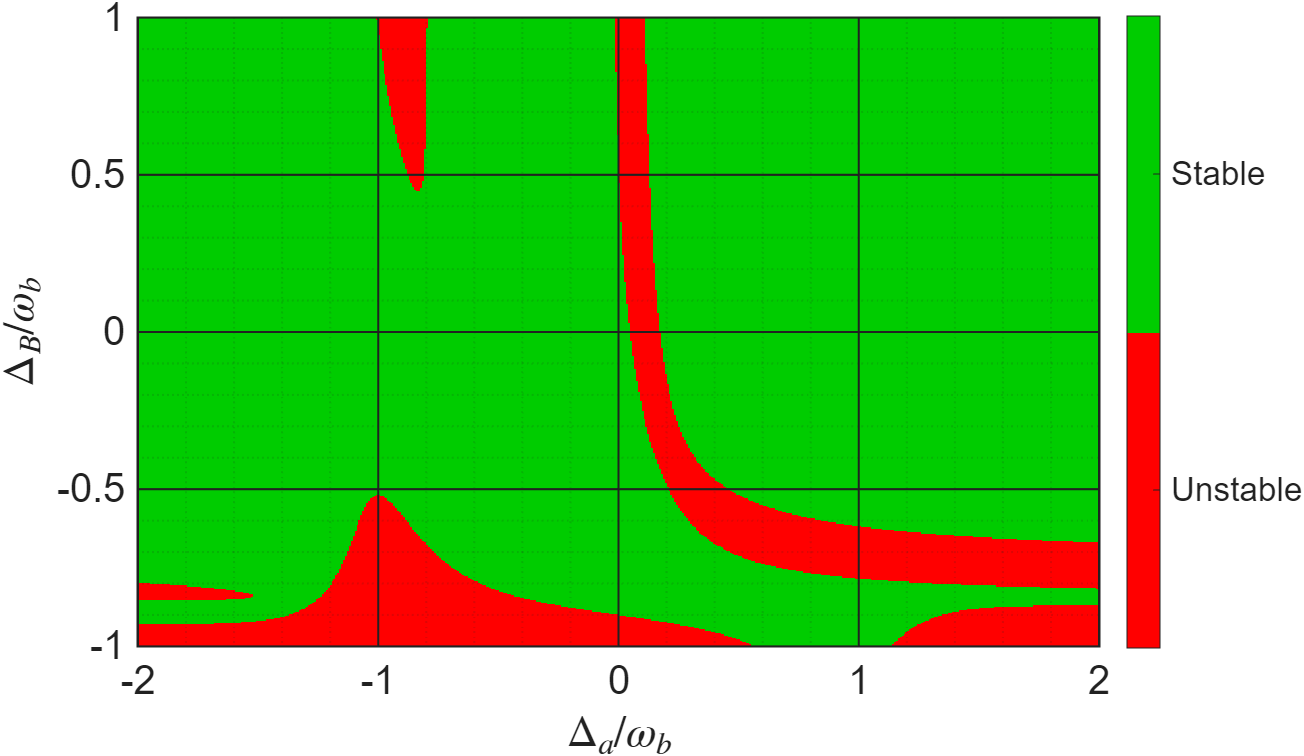}
		\put(-51,127){\large(a)} 
	\hfil
	\includegraphics[width=0.48\textwidth]{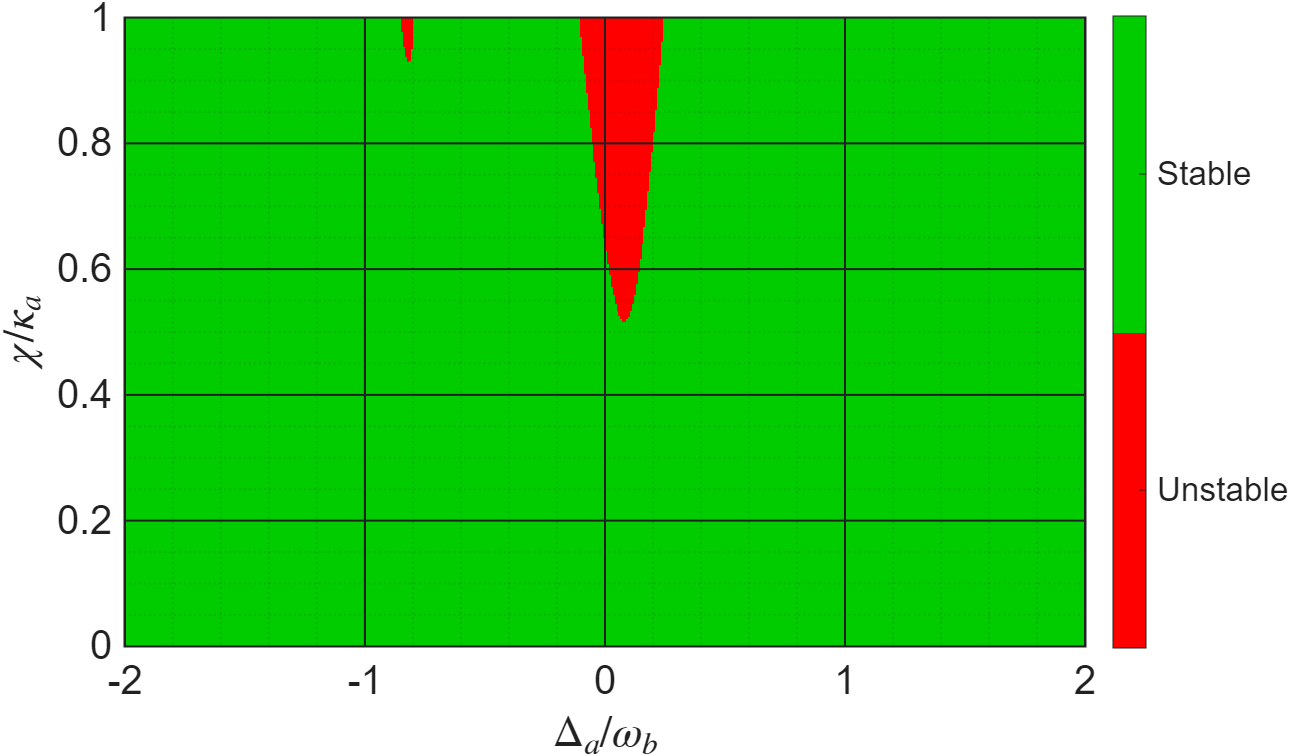}
			\put(-53,127){\large(b)}
	\hfil
	\includegraphics[width=0.48\textwidth]{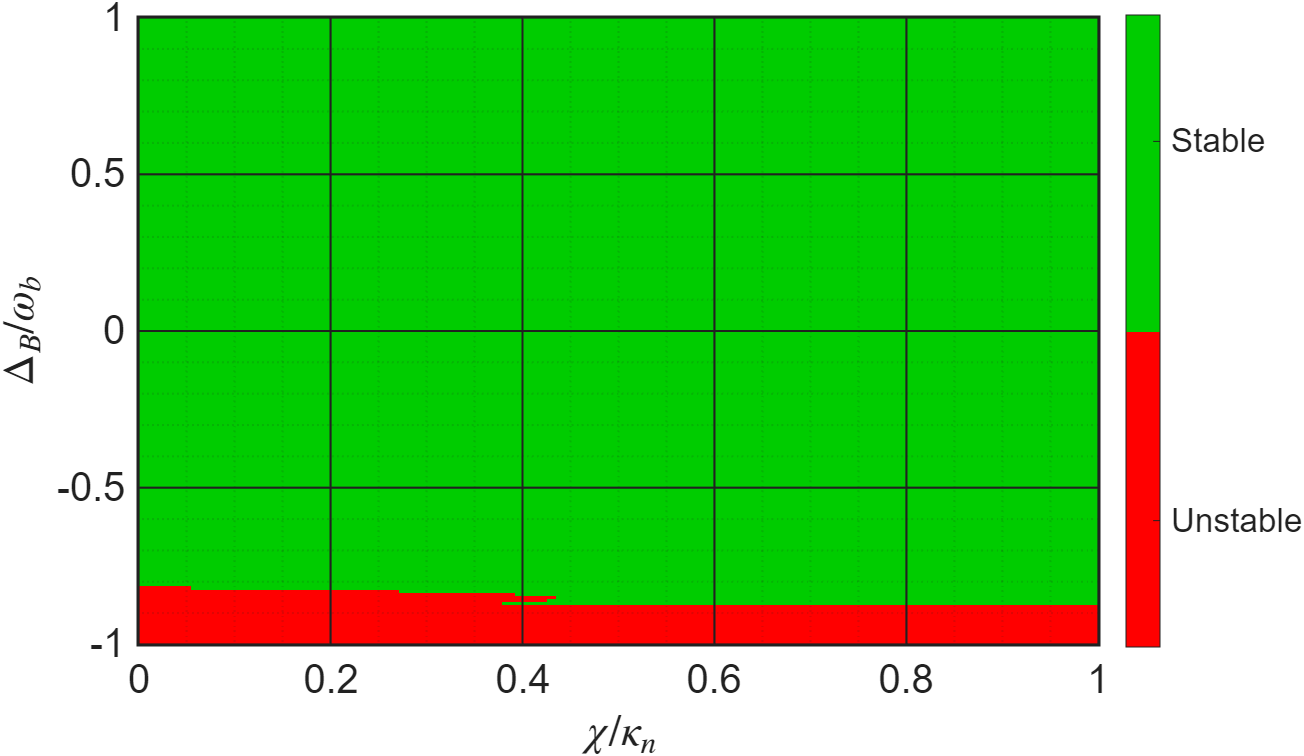}
			\put(-53,127){\large(c)}
	\caption{Plot of the stability condition versus $\Delta_B/\omega_b$ and the cavity detuning $\Delta_a$ (a), the nonlinear gain $\chi/\omega_b$ and the cavity detuning $\Delta_a$ (b), and the Barnett shift $\Delta_B/\omega_b$ and the nonlinear gain $\chi/\omega_b$ (c).} 
	\label{F10}
\end{figure}

To investigate the nonreciprocal bipartite and entanglement induced by the Barnett effect, we examine the stability of the CMM system through the Routh–Hurwitz criterion \cite{RC}. The system’s stability depends on the signs of the real parts of the eigenvalues associated with the dynamical evolution matrix. Determining these eigenvalues, denoted by $\Lambda$, requires solving the characteristic equation $|\mathcal{A} - \Lambda I| = 0$. As illustrated in Figure \ref{F10}, we delineate the stable and unstable parameter regimes. The parameters selected for this study are $-0.3 \leqslant \Delta_B / \omega_b \leqslant 0.3$ and $\chi = 0.6 \kappa_a$, both of which have been confirmed to reside within the stable region.

 \section{Conclusion}\label{sec4} 
 In this paper, we have proposed an efficient scheme for generating and controlling nonreciprocal quantum entanglement in cavity CMM systems by exploiting the Barnett effect and the presence of the OPA. With appropriate parameter choices, perfect nonreciprocity in the entangled states can be achieved. Due to the Barnett effect, giant nonreciprocal entanglement can emerge. All entanglements with ideal nonreciprocity can be achieved via tuning the photon frequency detuning, appropriately choosing the cavity-magnon coupling regime and nonlinear gain, and the phase shift of OPA. Interestingly, the amount of entanglement nonreciprocity and her resilience against magnon thermal occupation are remarkably enhanced by increasing the gain of OPA. That nonreciprocity can be significantly enhanced even at relatively high temperatures. By adjusting the cavity detuning, the GIP can evidently exhibit strong nonreciprocal behavior. We demonstrate a practical scheme for realizing nonreciprocal single-phonon devices, which may enable applications in quantum information processing and quantum communication. The proposed scheme offers a pathway toward the development of novel nonreciprocal devices that maintain robustness against thermal fluctuations. Furthermore, our results lay the foundation for engineering nonreciprocal quantum resources based on the Barnett effect in cavity magnomechanical (CMM) systems. Overall, our findings underscore the feasibility of achieving ideal nonreciprocity in such platforms.

\section*{Acknowledgments}

The authors extend their appreciation to the Deanship of Research and Graduate Studies at King Khalid University for funding this work through Large Research Project under grant number RGP2/172/46.


\begin{thebibliography}{99} 
	
\bibitem{1} D. Lachance-Quirion, Y. Tabuchi, A. Gloppe, K. Usami, and
	Y. Nakamura, 
	Appl. Phys. Express 12, 070101 (2019).
\bibitem{2} H. Yuan, Y. Cao, A. Kamra, R. A. Duine, and P. Yan, Phys. Rep. 965, 1 (2022).
\bibitem{3}Y. Li, W. Zhang, V. Tyberkevych, W.-K. Kwok, A. Hoffmann,
	and V. Novosad, J. Appl. Phys.
	128, 130902 (2020).
\bibitem{4} Y.-P. Wang and C.-M. Hu, J. Appl. Phys. 127, 130901
	(2020).
\bibitem{5} X. Zuo, Z.-Y. Fan, H. Qian, M.-S. Ding, H. Tan, H. Xiong, and J. Li, New J. Phys. 26, 031201 (2024).
\bibitem{6} Y. Tabuchi, S. Ishino, T. Ishikawa, R. Yamazaki, K. Usami,
and Y. Nakamura, Phys. Rev. Lett. 113,
083603 (2014).
\bibitem{7} H. Huebl, C.W. Zollitsch, J. Lotze, F. Hocke, M. Greifenstein,
A. Marx, R. Gross, and S. T. B. Goennenwein, Phys. Rev. Lett. 111, 127003 (2013).
\bibitem{19}  R.-C. Shen, Y.-P. Wang, J. Li, S.-Y. Zhu, G. S. Agarwal, and
J. Q. You, Phys. Rev. Lett. 127,
183202 (2021).
\bibitem{20}J. Li, S.-Y. Zhu, and G. S. Agarwal, Phys. Rev. Lett. 121,
203601 (2018).
\bibitem{24} D. Zhang, X.-Q. Luo, Y.-P. Wang, T.-F. Li, and J. Q. You,
Nat. Commun. 8, 1368 (2017).
\bibitem{32}T. Trickle, Z. Zhang, and K. M. Zurek, Phys. Rev. Lett. 124, 201801 (2020).
\bibitem{13} X. Zhang, C. L. Zou, L. Jiang, and H. X. Tang,
Sci. Adv. 2, e1501286 (2016).
\bibitem{36} Z.-Y. Fan, L. Qiu, S. Gröblacher, and J. Li, Laser
Photonics Rev. 17, 2200866 (2023).
\bibitem{39} C. Kong, B. Wang, Z.-X. Liu, H. Xiong, and Y. Wu,
Opt. Express 27, 5544 (2019).
\bibitem{41} M. F. Colombano, G. Arregui, F. Bonell, N. E. Capuj, E.
Chavez-Angel, A. Pitanti, S. O. Valenzuela, C.M. Sotomayor-
Torres, D. Navarro-Urrios, and M. V. Costache, Phys. Rev.
Lett. 125, 147201 (2020).
\bibitem{44} S.-F. Qi and J. Jing, Phys. Rev. A
103, 043704 (2021).

\bibitem{14} G.-T. Xu, M. Zhang, Y. Wang, Z. Shen, G.-C. Guo, and
C.-H. Dong, Phys. Rev. Lett. 131, 243601 (2023).

\bibitem{15}
R.-C. Shen, J. Li, Z.-Y. Fan, Y.-P. Wang, and J. Q. You,
Phys. Rev. Lett. 129, 123601 (2022).

\bibitem{16} Z. Shen, G.-T. Xu, M. Zhang, Y.-L. Zhang, Y. Wang, C.-Z.
Chai, C. L. Zou, G.-C. Guo, and C. H. Dong, Phys. Rev.
Lett. 129, 243601 (2022).
\bibitem{gip1}
 D. Girolami, A. M. Souza, V. Giovannetti, T. Tufarelli, J. G. Filgueiras, R. S. Sarthour and  G. Adesso . Physical Review Letters. 112, 210401 (2014).
 \bibitem{gip} G. Adesso,  Phys. Rev. A. 90, 022321 (2014).
 \bibitem{z17}
 C. Sabín, G. Adesso, Phys. Rev. A 92, 042107 (2015).
\bibitem{46}
C. Caloz, A. Alù, S. Tretyakov, D. Sounas, K. Achouri, and
Z.-L. Deck-Léger, Phys. Rev.
Appl. 10, 047001 (2018). 
\bibitem{47}E. Theveneau, B. Steventon, E. Scarpa, S. Garcia, X. Trepat,
A. Streit, and R. Mayor, Nat.
Cell Biol. 15, 763 (2013).
\bibitem{48} Y. Duan, J. Agudo-Canalejo, R. Golestanian, and B.
Mahault, Phys. Rev. Lett. 131, 148301 (2023).

\bibitem{49}
 D. L. Sounas and A. Alù, Nat. Photon. 11, 774 (2017).
 \bibitem{50}
 Q.-T. Cao, H. Wang, C.-H. Dong, H. Jing, R.-S. Liu, X. Chen,
 L. Ge, Q. Gong, and Y.-F. Xiao, Phys.
 Rev. Lett. 118, 033901 (2017).
 \bibitem{55} X. Lu, W. Cao, W. Yi, H. Shen, and Y. Xiao, Phys. Rev. Lett. 126, 223603 (2021).
 \bibitem{57} S. Manipatruni, J. T. Robinson, and M. Lipson, Phys. Rev. Lett. 102,
 213903 (2009).
 \bibitem{61} S. Maayani, R. Dahan, Y. Kligerman, E. Moses, A. U. Hassan,
 H. Jing, F. Nori, D. N. Christodoulides, and T. Carmon, Nature (London) 558, 569 (2018).
 \bibitem{70} Y. F. Jiao, S.-D. Zhang, Y.-L. Zhang, A. Miranowicz, L.-M.
 Kuang, and H. Jing, Phys. Rev. Lett. 125,
 143605 (2020).
 \bibitem{79} R. Horodecki, P. Horodecki, M. Horodecki, and K. Horodecki, Rev. Mod. Phys. 81, 865 (2009).
 \bibitem{81}
 F. Xu, X. Ma, Q. Zhang, H.-K. Lo, and J.-W. Pan, Rev. Mod.
 Phys. 92, 025002 (2020).
 \bibitem{z4}
S.  Ullah, H. S. Qureshi,  G. Tiaz, F. Ghafoor, and  F.  Saif,  . Applied Optics, 58, 197-204 (2018).
 \bibitem{z5}
H. S. Qureshi, S. Ullah, and  F. Ghafoor, Scientific reports. 8, 16288 (2018).
 \bibitem{84}
 X.-C. Yao, T.-X. Wang, P. Xu, H. Lu, G.-S. Pan, X.-H.
 Bao, C.-Z. Peng, C.-Y. Lu, Y.-A. Chen, and J. W. Pan, Nat. Photon. 6, 225
 (2012).
 \bibitem{86} X.-Y. Luo, Y. Yu, J.-L. Liu, M.-Y. Zheng, C.-Y. Wang, B.
 Wang, J. Li, X. Jiang, X.-P. Xie, Q. Zhang, X.-H. Bao, and
 J.-W. Pan, Phys. Rev. Lett. 129, 050503
 (2022).
 \bibitem{88} V. Krutyanskiy, M. Galli, V. Krcmarsky, S. Baier, D. A.
 Fioretto, Y. Pu, A. Mazloom, P. Sekatski, M. Canteri, M.
 Teller, J. Schupp, J. Bate, M. Meraner, N. Sangouard, B. P.
 Lanyon, and T. E. Northup, Phys. Rev. Lett. 130, 050803
 (2023).
 \bibitem{89} T. Palomaki, J. Teufel, R. Simmonds, and K. W. Lehnert, Science
 342, 710 (2013). 
 
 Kuang, and H. Jing, Phys. Rev. Lett. 125,
 143605 (2020).
 \bibitem{72} S. Chakraborty and C. Das, Phys. Rev.
 A 108, 063704 (2023).
 \bibitem{75} J. Chen, X.-G. Fan, W. Xiong, D. Wang, and L. Ye, Phys.
 Rev. B 108, 024105 (2023).
 \bibitem{77} T.-X. Lu, B. Li, Y. Wang, D.-Y. Wang, X. Xiao, and H.
 Jing, Phys. Rev. Appl. 22, 064001
 (2024).
  \bibitem{z2}
 M. Amazioug, S. Abdel‐Khalek,  M. Asjad, Annalen der Physik, e00289 (2025).
 \bibitem{z9}
 X. H. Fan, Y. N. Zhang, J. P. Yu, M. Y. Liu, W. D. He, H. C. Li, W. Xiong, Advanced Quantum Technologies. 7, 2400043 (2024).
 \bibitem{107} K.-W. Huang, X.Wang, Q.-Y. Qiu, and H. Xiong, Opt. Lett. 49, 758
 (2024).
 \bibitem{PRA}
 B. Hussain, S. Qamar, and  M. Irfan. Physical Review A. 105, 063704(2022).
  \bibitem{pla}  N. chabar , M. B. Amghar, and  M. Amazioug, Physics Letters A. 519, 129712 (2024).
  \bibitem{z7}
  M. Amghar,  N. Chabar, and  M. Amazioug, Journal of the Optical Society of America B. 42, 120-128 (2024).
 \bibitem{z3}
H. Harraf, N. Chabar, M. Amazioug, R. Ahl Laamara, S.  Haddadi,  Scientific Reports. 15, 36742 (2025).
 \bibitem{z1}
M. Amazioug, B. Teklu, and   M. Asjad, Scientific Reports. 13, 3833  (2023). 

\bibitem{z10}
A. Sohail, M. Amazioug, S. K. Singh, N. Chabar, R. Ahmed,  M. C. de Oliveira, Annalen der Physik, 2400375 (2025).

 \bibitem{97} A. Kani, F. Quijandría, and J. Twamley,
 Phys. Rev. Lett. 129, 257201 (2022).
 
 \bibitem{102} H. Chudo, M. Ono, K. Harii, M. Matsuo, J. Ieda, R. Haruki, S.
 \bibitem{105} A. Manjavacas and F. J. García de Abajo, Phys. Rev. Lett. 105, 113601 (2010).
 Okayasu, S. Maekawa, H. Yasuoka, and E. Saitoh,
 Appl. Phys. Express 7, 063004 (2014).
 \bibitem{104} C. S. Davies, F. G. N. Fennema, A. Tsukamoto, I. Razdolski,
 A. V. Kimel, and A. Kirilyuk, Nature (London) 628,
 540 (2024).
 
 \bibitem{106} M. Imai, H. Chudo, M. Ono, K. Harii, M. Matsuo, Y. Ohnuma, S. Maekawa, and E. Saitoh, Appl. Phys.
 Lett. 114, 162402 (2019).
 
 \bibitem{n1}
 N. Chabar,  M. Amazioug. arXiv preprint arXiv:2507.09590 (2025).
 \bibitem{H1}
  P. C. Ge, Y. Yu, H. T. Wu, X. Han, H. F. Wang, , S. Zhang,  Scientific Reports, 15, 7937 (2025).
 \bibitem{n2} Y. Liu, R. S. Zhao, K. K. Zhang, J. F. Li, R. G. Wan, H. Sun, and  X. T. Xie, Chaos, Solitons \& Fractals, 201, 117368(2025).
 \bibitem{108} D. Vitali, S. Gigan, A. Ferreira, H. R. Böhm, P. Tombesi,
 A. Guerreiro, V. Vedral, A. Zeilinger, and M. Aspelmeyer,
 Optomechanical entanglement between a movable mirror and
 a cavity field, Phys. Rev. Lett. 98, 030405 (2007).
 
 \bibitem{109}G. Vidal and R. F. Werner.
 Phys. Rev. A 65, 032314 (2002).
 
  \bibitem{110} M. B. Plenio, Phys. Rev. Lett. 95, 090503
 (2005).
 \bibitem{bs} Lu, T. X., Li, Z. S., Chen, L. S., Wang, Y., Xiao, X., and  H. Jing,  Physical Review A. 111, 013713 (2025).
 
 \bibitem{pr} M. Amazioug, B. Maroufi, and M. Daoud,  Quantum Information Processing. 19, 1-16  (2020).

 \bibitem{april2025}  M. Y. Liu, Y. Gong, J. Chen, Y. W. Wang, and  W. Xiong, Chinese Physics B. 34, 057202 (2025).

 \bibitem{RC} E. X. DeJesus and C. Kaufman, Phys. Rev. A 35, 5288 (1987).
\end{thebibliography}
\end{document}